\documentclass[10pt,twocolumn,journal]{IEEEtran}
\usepackage{latexsym}
\usepackage{float}
\usepackage{amsfonts}
\usepackage{amsbsy}
\usepackage{amssymb}
\usepackage{times}
\usepackage{graphicx}
\usepackage{enumerate}
\usepackage[usenames]{color}
\usepackage[dvips]{pstcol}
\usepackage{epstopdf}
\usepackage{cite}
\usepackage{amssymb}
\usepackage{amsfonts}
\usepackage{graphicx}
\usepackage{epsfig}
\usepackage{psfrag}
\usepackage{xcolor}
\usepackage{amsfonts, bm}
\usepackage{epstopdf}
\usepackage{cite}
\usepackage{color}
\usepackage{xcolor}
\usepackage{subfig}
\usepackage{verbatim}
\usepackage{multirow}
\usepackage{array}
\usepackage{booktabs}
\usepackage{amsthm}

\usepackage{algorithm}
\usepackage{algorithmicx}
\usepackage{algpseudocode}
\usepackage{amsmath}

\IEEEoverridecommandlockouts
\begin{document}
	\title{ Two-Timescale End-to-End Learning for Channel Acquisition and Hybrid Precoding }
	\author{ Qiyu Hu, \textit{Student Member, IEEE,} Yunlong Cai, \textit{Senior Member, IEEE,} Kai Kang, \\ Guanding Yu, \textit{Senior Member, IEEE,}  Jakob Hoydis, \textit{Senior Member, IEEE,}  and Yonina C. Eldar, \textit{Fellow, IEEE} 
		\thanks{
			
			Q. Hu, Y. Cai, K. Kang, and G. Yu are with the College of Information Science and Electronic Engineering, Zhejiang University, Hangzhou 310027, China (e-mail: qiyhu@zju.edu.cn; ylcai@zju.edu.cn; kangkai@zju.edu.cn; yuguanding@zju.edu.cn).
			
			J. Hoydis is with NVIDIA, 06906 Sophia Antipolis, France (e-mail: jhoydis@nvidia.com). 
			
			Y. C. Eldar is with the Department of Mathematics and Computer Science, Weizmann Institute of Science, Rehovot 7610001, Israel (e-mail: yonina.eldar@weizmann.ac.il).
		}
	}

\maketitle
\vspace{-3.3em}
\begin{abstract}
In this paper, we propose an end-to-end deep learning-based joint transceiver design algorithm for millimeter wave (mmWave) massive multiple-input multiple-output (MIMO) systems, which consists of deep neural network (DNN)-aided pilot training, channel feedback, and hybrid analog-digital (HAD) precoding.
Specifically, we develop a DNN architecture that maps the received pilots into feedback bits at the receiver, and then further maps the feedback bits into the hybrid precoder at the transmitter.
To reduce the signaling overhead and channel state information (CSI) mismatch caused by the transmission delay, a two-timescale DNN composed of a long-term DNN and a short-term DNN is developed. The analog precoders are designed by the long-term DNN based on the CSI statistics and updated once in a frame consisting of a number of time slots. In contrast, the digital precoders are optimized by the short-term DNN at each time slot based on the estimated low-dimensional equivalent CSI matrices. 
A two-timescale training method is also developed for the proposed DNN with a binary layer. We then analyze the generalization ability and signaling overhead for the proposed DNN based algorithm. Simulation results show that our proposed technique significantly outperforms conventional schemes in terms of bit-error rate performance with reduced signaling overhead and shorter pilot sequences.
\end{abstract}
\begin{IEEEkeywords}
Deep learning, massive multiple-input multiple-output (MIMO), millimeter wave, hybrid precoding, channel acquisition, two-timescale.
\end{IEEEkeywords}

\IEEEpeerreviewmaketitle

\section{Introduction}
Millimeter wave (mmWave) communications have been recognized as one of the key technologies to meet the requirement of high data rate transmission in the development of 5G wireless networks due to the enormous bandwidth \cite{mmWavemag}. The short wavelength of mmWave makes it feasible to utilize large-scale antenna arrays, where massive multiple-input multiple-output (MIMO) systems provide adequately large array gains for spatial multiplexing, hence improving system capacity and alleviating radio spectrum shortage \cite{MIMOoverview, MIMO01, MIMO02}. However, conventional fully-digital precoding leads to expensive fabrication costs and high energy consumption, which are the main obstacles for equipping a radio frequency (RF) chain for each antenna. To address this problem, hybrid analog-digital (HAD) precoding has been suggested, where a large number of antennas are connected to fewer RF chains \cite{Hybrid01, Spatially, Near, PDD}. Channel estimation, feedback, and hybrid precoding design for HAD system is a challenging and growing research area. Here we propose a deep learning-based approach for jointly designing these modules in an efficient manner. 

\subsection{Prior Work}
Conventional MIMO communication systems with HAD architectures are typically designed as follows \cite{mmWavemag, MIMOoverview, MIMO01, MIMO02}. The transmitter (TX) first sends pilots to the receiver (RX), which applies a sparse recovery algorithm to estimate channel parameters in the angular domain \cite{ChannelEsi00, ChannelEsi01, OMP}. Then, the RX quantizes the estimated channel state information (CSI) and feeds back these quantized channel parameters to the TX \cite{ChannelFeed01, ChannelFeed02, ChannelFeed11, ChannelFeed12}. Subsequently, the TX receives the quantized parameters and uses them to estimate the CSI. The precoding matrices are then designed based on the recovered CSI \cite{HybridLarge, Alternate, Limited, Codebook, MaGiQ, ChannelMatch}. 
Conventional schemes of channel estimation, feedback, and hybrid precoding are designed separately due to the intractability of joint optimization.

Current feedback schemes in the literature can be mainly classified into two types: (i) exploiting spatial or temporal correlation of CSI to reduce the feedback overhead \cite{ChannelFeed01, ChannelFeed02}, and (ii) codebook-based schemes \cite{ChannelFeed11, ChannelFeed12}.
For precoding, the authors in \cite{HybridLarge} demonstrated that hybrid precoding with twice as many RF chains as data streams approaches the performance of fully-digital precoding. 
A manifold-based iterative algorithm for hybrid precoder design is developed in \cite{Alternate}. A hybrid precoding algorithm employing the sparse channel characteristics of multi-user mmWave systems is considered in \cite{Limited}, while \cite{Codebook} introduces a codebook-based hybrid precoding algorithm. The authors in \cite{MaGiQ} develop a family of algorithms that approximate the optimal fully-digital precoder with a hybrid one.  	
These conventional approaches generally achieve good performance for systems with (i) sufficiently large pilot length for channel estimation and (ii) a large amount of feedback bits in which the quantization error is small. 	

In contrast to conventional communication system designs that develop each module separately, an end-to-end deep learning framework is suitable to jointly design these modules \cite{DLMaga}. Deep learning can achieve satisfactory performance with reduced pilot length and smaller number of feedback bits \cite{FeedbackFDD}. Furthermore, it implicitly learns the CSI distributions in a data-driven manner from the process of optimizing the end-to-end communication system, without requiring precise mathematical CSI models \cite{DLMaga}. Finally, the computation of deep neural networks (DNNs) can be parallelized and its computational complexity can be much lower than that of conventional algorithms. 

Recently, deep learning has received considerable attention in communication systems \cite{FeedbackFDD, DLMaga, BeamSelect, DNNSystem, Eldar01, Eldar02, DNNesti01, DNNesti02, DNNFeed01, DNNFeed02, LearnOpt, Qiyu,HybridDNN, DNNHybrid01, DNNHybrid02}. Channel estimation and digital precoders are jointly designed in \cite{FeedbackFDD} for a frequency-division duplex (FDD) system to maximize the sum-rate. In \cite{BeamSelect}, beam selection and precoding are also jointly designed. 
Deep learning has further been applied to symbol detection \cite{Eldar01, Eldar02}, channel estimation \cite{DNNesti01, DNNesti02}, channel feedback \cite{DNNFeed01, DNNFeed02}, and precoding \cite{LearnOpt, Qiyu, HybridDNN, DNNHybrid01, DNNHybrid02}.
Specifically, in \cite{DNNesti01} and \cite{DNNesti02}, channel correlation and statistics are exploited by the DNN to improve the accuracy of channel estimation. The authors in \cite{DNNFeed01} and \cite{DNNFeed02} employed deep learning to solve the CSI feedback and reconstruction problem at the TX, by assuming that perfect CSI is available at the RX. In \cite{LearnOpt} and \cite{Qiyu}, black-box DNNs and model-driven DNNs have been respectively applied to optimize digital precoders. In \cite{HybridDNN, DNNHybrid01, DNNHybrid02}, an autoencoder-like DNN is employed to design hybrid precoding matrices.  

The aforementioned hybrid precoding algorithms are developed based on the instantaneous CSI. In the scenario of large-scale antennas, the acquisition of high-dimensional CSI matrices leads to heavy signaling overhead, which causes serious transmission delay and CSI mismatch. In \cite{SSCA, TwotimePrecode, TwotimeChannel}, a number of two-timescale hybrid precoding algorithms have been proposed, where the long-term analog precoders are optimized based on the CSI statistics and the short-term digital precoders are designed by employing the low-dimensional real-time equivalent CSI matrices. These two-timescale algorithms can reduce the signaling overhead and hence increase robustness against CSI errors caused by transmission delay. However, the existing two-timescale algorithms generally have high computational complexity and cannot jointly design the modules of channel estimation, feedback, and precoding in communication systems \cite{SSCA}.

\subsection{Motivation and Contribution}
The design of hybrid precoding matrices, channel estimation and feedback are challenging due to the constant modulus constraints of the analog precoder and the high dimension of the channel matrix. In addition, joint transceiver design still remains an open issue.  
To address these problems, we propose an end-to-end deep learning-based  joint transceiver design algorithm that encapsulates all modules of an FDD massive MIMO system to minimize the bit-error rate (BER). 
Our approach consists of a DNN-based channel estimation, CSI quantization and feedback at the RX, and pilot design and hybrid precoding at the TX, where these DNNs are jointly trained. 
Specifically, we develop a DNN architecture that maps the received pilots into feedback bits at the RX, and then maps the feedback bits into the hybrid precoder at the TX. We model the feedback bits in the proposed DNN architecture as the outputs of binary neurons. In order to enable gradient-based training, we approximate the gradients of the binary layer with a variant of the straight-through (ST) estimator \cite{STestimator}.

To reduce CSI mismatch caused by the transmission delay and the heavy signaling overhead for CSI feedback due to the high dimension of the CSI matrix, we propose a two-timescale DNN composed of a long-term DNN and a short-term DNN. The time axis is partitioned into a sequence of superframes. We focus on a superframe that defines the long-timescale, during which the CSI statistics are assumed to stay nearly constant \cite{SSCA, TwotimePrecode, TwotimeChannel}. Each superframe is in turn partitioned into a sequence of frames. A frame contains a fixed number of time slots that define the short-timescale, during which the instantaneous CSI remains unchanged. Within each superframe, the long-term analog precoder and combiner are updated in a frame-based manner relying on the CSI statistics. The short-term digital precoder and combiner are optimized based on the low-dimensional real-time equivalent CSI within each time slot. 
Specifically, the TX sends low-dimensional pilots and the RX estimates the low-dimensional equivalent CSI matrix and feeds the quantized information back to the TX for the design of the digital precoder and combiner. The high-dimensional full CSI is estimated and fed back to update the analog precoder and combiner only once in a frame by the long-term DNN. 

We further develop a two-timescale training method for the proposed DNN and analyze the signaling overhead. We then consider techniques to improve the generalization ability of the proposed DNN. Generally, changes in the system parameters can be categorized into two types: (i) changes to the input distribution of the DNN, e.g., signal-to-noise ratio (SNR), and (ii) changes to the dimensions of some layers in the DNN, e.g., the number of feedback bits. For the former, we train the DNN based on a wider range of system parameters. For the latter, we modify the DNN and propose a two-step training method to enhance the generalization ability. 
The proposed two-timescale DNN can be easily extended to orthogonal frequency division multiplexing (OFDM) systems by simply modifying the structure of the training data. 

The main contributions of this paper are summarized as follows.
\begin{itemize}
\item We propose an end-to-end learning method for FDD mmWave MIMO systems, which includes channel estimation, quantization, feedback, and hybrid precoding.

\item A two-timescale DNN composed of a long-term DNN and a short-term DNN is developed to reduce the signaling overhead and CSI mismatch caused by the transmission delay. 

\item A two-timescale training method is also developed for the proposed DNN with a binary layer.  

\item Simulation results show that our proposed algorithm significantly outperforms conventional schemes in terms of BER performance with reduced pilot length and signaling overhead. 
\end{itemize}

\begin{figure*}[t]
\begin{centering}
\includegraphics[width=0.8\textwidth]{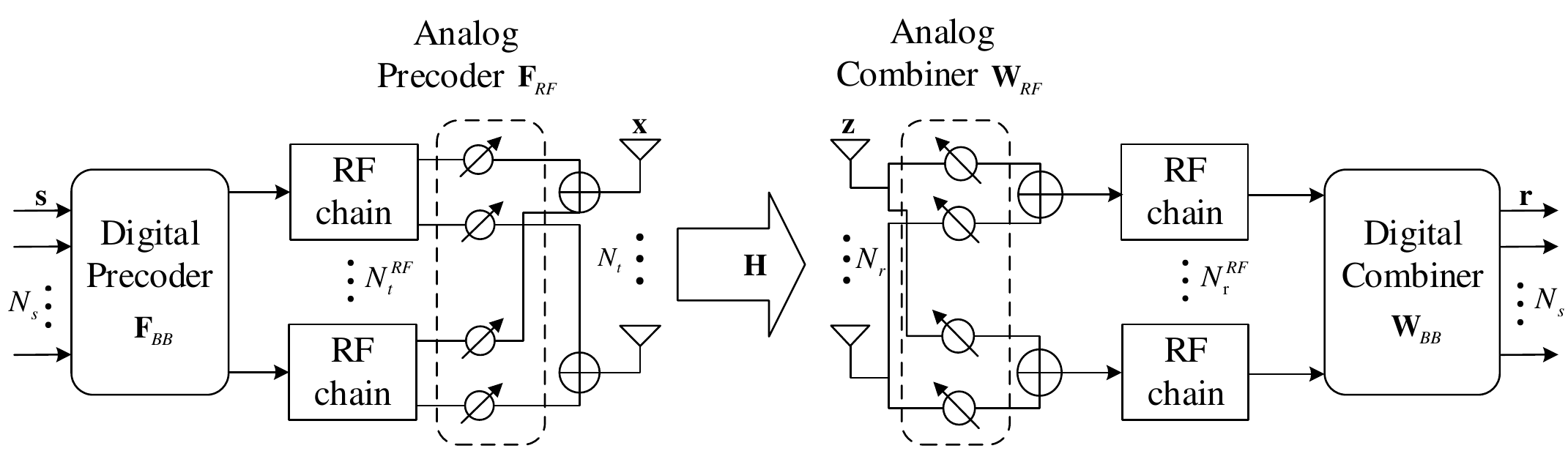}
\par\end{centering}
\caption{End-to-end mmWave FDD MIMO system with hybrid processing architecture.}
\label{HybridSystem}
\end{figure*}

\subsection{Organization and Notation}
The rest of the paper is structured as follows. Section \ref{System} introduces the system model and formulates our problem mathematically. Section \ref{DNNdesign} develops a deep learning framework for the investigated system and proposes a two-timescale DNN based on this framework. We present the implementation details and develop a training method for the proposed two-timescale DNN in Section \ref{Impleme}. In Section \ref{Analysis}, we develop the method for enhancing the generalization ability of the proposed DNN and analyze the signaling overhead. We present simulation results in Section \ref{Simulation} and conclude the paper in Section \ref{Conclusion}.

\emph{Notation:} Scalars, vectors, and matrices are respectively denoted by lower case, boldface lower case, and boldface upper case letters.
The notation $\mathbf{I}$ represents an identity matrix and $\mathbf{0}$ is an all-zero matrix.
For a matrix $\mathbf{A}$, ${\bf{A}}^T$, $\mathbf{A}^*$, ${\bf{A}}^H$, and $\|\mathbf{A}\|$ are its transpose, conjugate, conjugate transpose, and Frobenius norm, respectively.
For a vector $\mathbf{a}$, $\|\mathbf{a}\|$ is its Euclidean norm. We use $\mathbb{E}\{ \cdot \}$ for the statistical expectation,
$\Re\{ \cdot \}$ ($\Im\{ \cdot \}$) denotes the real (imaginary) part of a variable, $\textrm{Tr}\{ \cdot \}$ denotes the trace operation, $| \cdot |$ is the absolute value of a complex scalar, and $\circ$ is the element-wise multiplication of two matrices, i.e., Hadmard product.
Finally, ${\mathbb{C}^{m \times n}}\;({\mathbb{R}^{m \times n}})$ are the space of ${m \times n}$ complex (real) matrices.

\section{System Model and Problem Formulation} \label{System}

In this section, we introduce the end-to-end mmWave MIMO system model and then formulate our problem mathematically.

\subsection{End-to-End mmWave MIMO System}
\subsubsection{Signal Model}
Consider an end-to-end mmWave FDD MIMO system, where a TX equipped with $N_{t}$ transmit antennas and $N^{RF}_{t}$ RF chains sends $N_{s}$ data streams, $N_{s}\leq N^{RF}_{t} \leq N_{t}$, to a RX equipped with $N_r$ receive antennas and $N^{RF}_{r}$ RF chains, $N_{s}\leq N^{RF}_{r} \leq N_{r}$. At the TX, the RF chains are followed by a network of phase shifters that expands the $N^{RF}_{t}$ digital outputs to $N_{t}$ precoded analog signals feeding the transmit antennas. Similarly, at the RX, the $N_{r}$ receive antennas are followed by a network of phase shifters that feed the $N^{RF}_{r}$ RF chains.

The TX transmits $N_s$ parallel data streams $\mathbf{S}_{b} \in \{ 0,1 \}^{N_{s} \times \log_{2} M}$, which consist of binary bits with dimension $N_{s}\times \log_{2}M$. They are mapped into the symbol $\mathbf{s}\in \mathbb{C}^{N_s \times 1}$ according to an $M$-ary modulation scheme, where we assume $\mathbb{E}\{\mathbf{s}\mathbf{s}^H\} = \mathbf{I}_{N_{s}}$. As illustrated in Fig. \ref{HybridSystem}, the symbol vector $\mathbf{s}$ is processed through a digital precoder $\mathbf{F}_{BB}\in \mathbb{C}^{N^{RF}_{t} \times N_s}$, and then an analog precoder $\mathbf{F}_{RF}\in \mathbb{C}^{N_t \times N^{RF}_{t}}$. The precoded signal vector $\mathbf{x}\in \mathbb{C}^{N_t \times 1}$ can be written as
\begin{equation}
\mathbf{x} = \mathbf{F}_{RF}\mathbf{F}_{BB}\mathbf{s}. 
\end{equation}
Here $\mathbf{F}_{RF}$ denotes a phase-only modulation by phase shifters, which follows the constant modulus constraint $ | [\mathbf{F}_{RF}]_{mn} | = 1, \forall m, n$. The matrix $\mathbf{F}_{BB}$ is normalized such that $\| \mathbf{F}_{RF}\mathbf{F}_{BB} \|_{F}^{2} = P_{T} $ to meet the power constraint at the TX, where $P_{T}$ denotes the maximum transmission power. The precoded signal $\mathbf{x}$ is transmitted over a narrowband block-fading propagation channel. The received analog signal vector $\mathbf{z}\in \mathbb{C}^{N_r \times 1}$ at the RX's antennas is given by
\begin{equation}
\mathbf{z} = \mathbf{H}\mathbf{F}_{RF}\mathbf{F}_{BB}\mathbf{s}+\mathbf{n}, 
\end{equation}
where $\mathbf{H}\in \mathbb{C}^{N_{r}\times N_{t}}$ denotes the channel matrix and $\mathbf{n}\sim \mathcal{CN}(\mathbf{0}, \sigma_{n}^2 \mathbf{I}_{N_{r}})$ is additive white Gaussian noise (AWGN).

Similar to the design of hybrid precoders, an analog combiner $\mathbf{W}_{RF}\in \mathbb{C}^{N_{r}\times N^{RF}_{r}}$ is employed at the RX, followed by a digital baseband combiner $\mathbf{W}_{BB} \in \mathbb{C}^{N^{RF}_{r} \times N_s}$. The detected signal is written as
\begin{equation} \label{SignalModel}
\mathbf{r} = \mathbf{W}^{H}_{BB}\mathbf{W}^{H}_{RF}\mathbf{H}\mathbf{F}_{RF}\mathbf{F}_{BB}\mathbf{s} + \mathbf{W}^{H}_{BB}\mathbf{W}^{H}_{RF}\mathbf{n},
\end{equation}
where $\mathbf{W}_{RF}$ meets the hardware constraint $| [\mathbf{W}_{RF}]_{rs} | = 1, \forall r, s$. Finally, the detected signal vector $\mathbf{r}$ is demodulated to recover the original bits of the $N_s$ data streams, and yields the estimated $\hat{\mathbf{S}}_{b}$.

\begin{figure*}[t]
\begin{centering}
\includegraphics[width=0.95\textwidth]{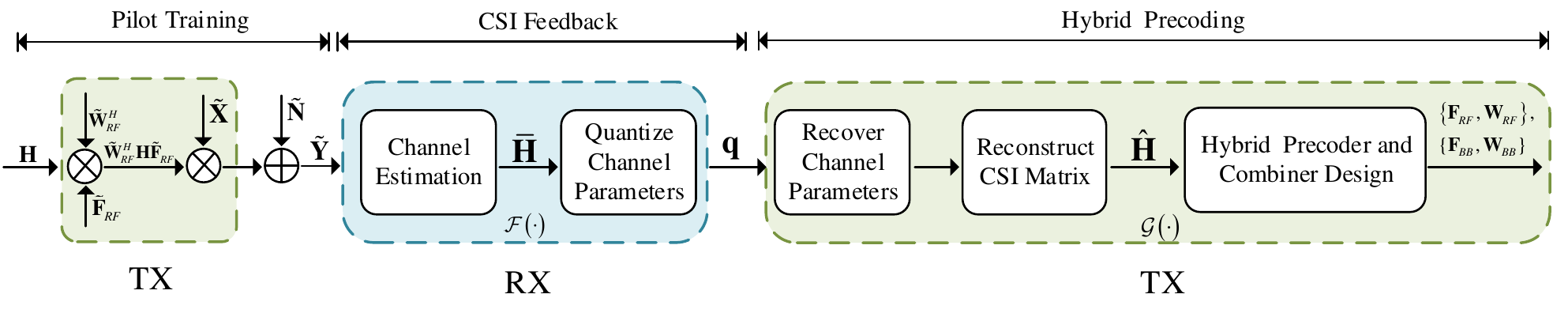}
\par\end{centering}
\caption{The conventional scheme where the pilot training and hybrid precoding are executed at the TX and the CSI feedback is executed at the RX.}
\label{FrameCSI}
\end{figure*}

In the following, we present the detailed communication process in Fig. \ref{FrameCSI}, which consists of channel estimation, feedback, and hybrid precoding. 

\subsubsection{Pilot Training for Channel Estimation and CSI Feedback}

It is important for the TX to acquire the CSI matrix $\mathbf{H}$ for hybrid precoding. It is assumed that the TX and RX have no prior knowledge of the CSI and it is estimated by pilot training. In particular, we consider a pilot training stage, prior to the data transmission stage. The TX sends training pilots $\tilde{\mathbf{X}}\in \mathbb{C}^{N_{t}^{RF} \times L}$ with length $L$, and the RX receives $\tilde{\mathbf{Y}}\in \mathbb{C}^{N_{r}^{RF} \times L}$ as
\begin{equation}
\tilde{\mathbf{Y}} = \tilde{\mathbf{W}}_{RF}^{H} \mathbf{H}\tilde{\mathbf{F}}_{RF}\tilde{\mathbf{X}} + \tilde{\mathbf{N}}, \label{signal}
\end{equation}
where $\tilde{\mathbf{F}}_{RF}\in \mathbb{C}^{N_t \times N^{RF}_{t}}$ and $\tilde{\mathbf{W}}_{RF}\in \mathbb{C}^{N_r \times N^{RF}_{r}}$ represent the analog precoder and combiner in the pilot training stage, respectively, whose columns can be selected from the DFT matrix \cite{ChannelEsi00}. In addition, $\mathbf{N}\in \mathbb{C}^{N_{r} \times L}$ denotes the AWGN matrix, and $\tilde{\mathbf{N}} = \tilde{\mathbf{W}}_{RF}^{H}\mathbf{N}$. Note that the transmitted pilots in the $l$-th pilot transmission $\tilde{\mathbf{x}}_{l}$ (the $l$-th column of $\tilde{\mathbf{X}}$) should satisfy the power constraint, i.e., $\|\tilde{\mathbf{x}}_{l} \|^{2}\leq P$. 

The RX estimates the CSI matrix  $\mathbf{H}$ from the received signal $\tilde{\mathbf{Y}}$ \cite{ChannelEsi00, ChannelEsi01, OMP, ChannelFeed01}. It then extracts useful information, e.g., complex gain, azimuth angles-of-arrival (AoAs), and angles-of-departure (AoDs) of CSI, and subsequently feeds it back to the TX in the form of $B$ information bits as
\begin{equation}
\mathbf{q} = \mathcal{F}(\tilde{\mathbf{Y}}),
\end{equation}
where the mapping $\mathcal{F}: \mathbb{C}^{N_{r}^{RF} \times L} \rightarrow  \{\pm 1\}^{B}$ denotes the feedback scheme.

\subsubsection{Hybrid Precoding}
The TX collects the feedback bits $\mathbf{q}$ from the RX, and recovers the CSI parameters, e.g., AoAs and AoDs, to reconstruct the CSI matrix $\hat{\mathbf{H}}$ \cite{ChannelEsi01, ChannelFeed12}. Subsequently, the TX designs the hybrid precoders $\{\mathbf{F}_{RF}, \mathbf{F}_{BB}\}$ and combiners $\{\mathbf{W}_{RF}, \mathbf{W}_{BB}\}$ based on the reconstructed $\hat{\mathbf{H}}$ \cite{HybridLarge, Alternate, Limited, Codebook, MaGiQ}. The CSI reconstruction and hybrid precoding scheme at the TX are formulated as  
\begin{equation}
\{\mathbf{F}_{RF}, \mathbf{F}_{BB},\mathbf{W}_{RF}, \mathbf{W}_{BB}\} = \mathcal{G}(\mathbf{q}),
\end{equation}
where the TX receives $\mathbf{q}$ and maps it into the hybrid precoders and combiners, i.e., $\mathcal{G}(\mathbf{q})$.

\subsection{Two-timescale Frame Structure}
\begin{figure}[t]
\begin{centering}
\includegraphics[width=0.47\textwidth]{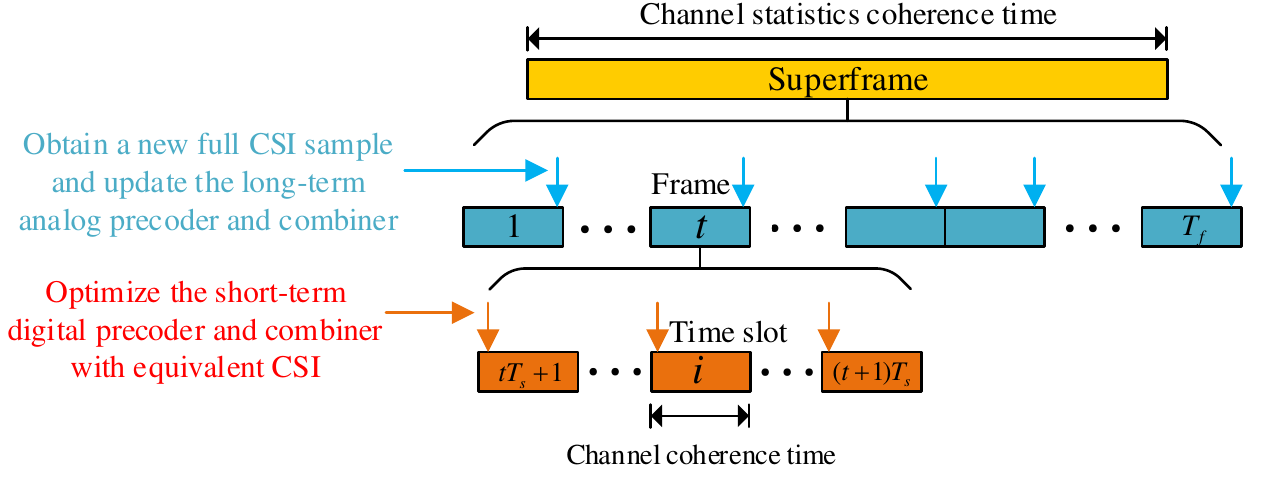}
\par\end{centering}
\caption{The frame structure of two-timescale hybrid precoding.}
\label{TwoTime}
\end{figure}

The joint design of the hybrid precoder and combiner for each instantaneous CSI is not realistic since it requires a large amount of overhead due to the estimation and feedback of high dimensional real-time CSI. It also requires extremely high computational complexity and hardware cost.  
To address these issues, we propose a two-timescale scheme that considers both the real-time equivalent CSI and channel statistics. As presented in Fig. \ref{TwoTime}, we focus on a particular superframe that is sufficiently large, during which the CSI statistics are assumed to be constant. It consists of $T_{f}$ frames, each of which is further divided into $T_{s}$ time slots and the instantaneous CSI keeps invariant within each time slot. Based on this partition, we define the following concepts of timescales:
\begin{itemize}
	\item Long-timescale: The CSI statistics are assumed to be constant over each superframe that consists of $T_{f}$ frames;
	\item Short-timescale: The instantaneous CSI is assumed invariant during each time slot.
\end{itemize}

Generally, the equivalent CSI matrix $\mathbf{H}_{eq}=\mathbf{W}_{RF}^{H}\mathbf{H}\mathbf{F}_{RF} \in \mathbb{C}^{N^{RF}_{r}\times N^{RF}_{t}}$ has much lower dimension than the full CSI matrix $\mathbf{H}\in \mathbb{C}^{N_{r}\times N_{t}}$. Thus, it is possible to obtain the real-time equivalent CSI matrix $\mathbf{H}_{eq}$ at each time slot by sending pilots. However, we can only acquire an outdated full CSI sample $\mathbf{H}$ at each frame since acquiring the real-time full CSI matrix $\mathbf{H}$ at each time slot will cause unacceptable signaling overhead in the massive MIMO scenario.  
Therefore, it is assumed that the RX is able to acquire a full CSI sample over each frame and it can acquire the real-time low-dimensional equivalent CSI matrix at each time slot.
In this way, we cannot optimize both the analog $\{\mathbf{F}_{RF}, \mathbf{W}_{RF}\}$ and digital $\{\mathbf{F}_{BB}, \mathbf{W}_{BB}\}$ based on $\mathbf{H}$ at each time slot. Thus, $\{\mathbf{F}_{RF}, \mathbf{W}_{RF}\}$ and $\{\mathbf{F}_{BB}, \mathbf{W}_{BB}\}$ have to be optimized at different timescale based on the outdated full CSI sample $\mathbf{H}$ and real-time equivalent CSI matrix $\mathbf{H}_{eq}$, respectively.
As shown in Fig. \ref{TwoTime}, the long-term analog precoder $\mathbf{F}_{RF}$ and combiner $\mathbf{W}_{RF}$ are updated at the end of each frame based on an estimated full CSI sample $\hat{\mathbf{H}}$ to achieve the massive MIMO array gain. In comparison, the short-term digital precoder $\mathbf{F}_{BB}$ and combiner $\mathbf{W}_{BB}$ are optimized in each time slot based on the estimated low-dimensional equivalent CSI matrix $\hat{\mathbf{H}}_{eq}$ to achieve the spatial multiplexing gain, while the long-term analog $\{\mathbf{F}_{RF}, \mathbf{W}_{RF}\}$ are fixed at these time slots.

\subsection{Problem Formulation}
The two-timescale problem of joint channel estimation, feedback, and hybrid precoding design can be formulated as
\begin{subequations}  \label{Problem}
\begin{eqnarray}
& \min\limits_{ \mathcal{X} } & \sum\limits_{t,i} P_{e}(\mathbf{F}^{t}_{RF}, \mathbf{F}^{i}_{BB}, \mathbf{W}^{t}_{RF}, \mathbf{W}^{i}_{BB})  \label{objBER}  \\
&\text{s.t.}  & | [\tilde{\mathbf{F}}_{RF}]_{mn} | = 1, \forall m, n,   \label{modulus11} \\
& & | [\tilde{\mathbf{W}}_{RF}]_{rs} | = 1, \forall r, s,  \label{modulus12} \\
& & | [\mathbf{F}_{RF}^{t}]_{mn} | = 1, \forall m, n, t, \label{modulus21}  \\
& & | [\mathbf{W}_{RF}^{t}]_{rs} | = 1,  \forall r, s, t, \label{modulus22} \\
& & \|\tilde{\mathbf{x}}_{l} \|^{2}\leq P, \forall l, \label{Pilots} \\
& & \|\tilde{\mathbf{x}}_{eq,l} \|^{2}\leq P, \forall l, \label{Pilotseq} \\
& & \| \mathbf{F}^{t}_{RF}\mathbf{F}^{i}_{BB} \|_{F}^{2} = P_{T}, \forall i, t, \label{Power} \\
& & \mathbf{q}^{t} = \mathcal{F}(\tilde{\mathbf{W}}_{RF}^{H} \mathbf{H}^{t} \tilde{\mathbf{F}}_{RF}\tilde{\mathbf{X}} + \tilde{\mathbf{N}}^{t}), \forall t, \label{FeedbackBit} \\
& & \mathbf{q}_{eq}^{i} = \mathcal{F}_{eq}( \mathbf{H}_{eq}^{i} \tilde{\mathbf{X}}_{eq} + \tilde{\mathbf{N}}_{eq}^{i}), \forall i, \label{FeedbackBiteq} \\
& & \{\mathbf{F}_{RF}^{t}, \mathbf{W}_{RF}^{t}\} = \mathcal{G}(\mathbf{q}^{t}), \forall t, \label{Precoder}  \\
& & \{ \mathbf{F}_{BB}^{i}, \mathbf{W}_{BB}^{i}\} = \mathcal{G}_{eq}(\mathbf{q}_{eq}^{i}), \forall i, \label{Precodereq}
\end{eqnarray}
\end{subequations}
where $\mathcal{X}\triangleq \{ \mathbf{F}_{RF}^{t}, \mathbf{F}_{BB}^{i}, \mathbf{W}_{RF}^{t}, \mathbf{W}_{BB}^{i}, \tilde{\mathbf{X}}, \tilde{\mathbf{X}}_{eq}, \tilde{\mathbf{W}}_{RF}, \tilde{\mathbf{F}}_{RF}, \\ \mathbf{q}^{t}, \mathbf{q}_{eq}^{i}, \mathcal{F}(\cdot), \mathcal{F}_{eq}(\cdot), \mathcal{G}(\cdot), \mathcal{G}_{eq}(\cdot), \forall t, i \}$. In particular, $\{ \tilde{\mathbf{X}}, \tilde{\mathbf{W}}_{RF}, \tilde{\mathbf{F}}_{RF} \}$ and $\tilde{\mathbf{X}}_{eq}$ denote the pilots for the estimation of full CSI $\mathbf{H}$ and equivalent CSI $\mathbf{H}_{eq}$, $\tilde{\mathbf{x}}_{l}$ and $\tilde{\mathbf{x}}_{eq,l}$ represent the $l$-th column of $\tilde{\mathbf{X}}$ and $\tilde{\mathbf{X}}_{eq}$, respectively, $\mathbf{q}^{t}$ are the feedback bits for $\mathbf{H}^{t}$ at the $t$-th frame and  $\mathbf{q}_{eq}^{i}$ denote the feedback bits for $\mathbf{H}_{eq}^{i}$ at the $i$-th time slot. In addition, $\mathbf{F}_{RF}^{t}$ and $\mathbf{W}_{RF}^{t}$ are the analog precoder and combiner at the $t$-th frame, $\mathbf{F}_{BB}^{i}$ and $\mathbf{W}_{BB}^{i}$ are the digital precoder and combiner at the $i$-th time slot, $\mathcal{F}(\cdot)$ and $\mathcal{F}_{eq}(\cdot)$ represent the CSI feedback schemes for $\mathbf{H}$ and $\mathbf{H}_{eq}$, $\mathcal{G}(\cdot)$ and $\mathcal{G}_{eq}(\cdot)$ denote the analog and digital pecoding schemes, respectively. These variables and schemes are designed to improve the BER performance $P_{e}(\cdot)$, which is an intricate non-linear function of $\{\mathbf{F}_{RF}^{t}, \mathbf{F}_{BB}^{i}, \mathbf{W}_{RF}^{t}, \mathbf{W}_{BB}^{i}\}$. 
The constraints \eqref{modulus11}-\eqref{modulus12} and \eqref{modulus21}-\eqref{modulus22} denote the constant modulus constraints for pilots and precoder/combiner, respectively. The constraints \eqref{Pilots}-\eqref{Pilotseq} and \eqref{Power} are the transmit power constraints for digital pilots and precoder, respectively. For clarity, we omit the indices $t$ and $i$ in the following sections.

\section{Proposed DNN for End-to-End Learning} \label{DNNdesign}

\begin{figure*}[t]
\begin{centering}
\includegraphics[width=0.95\textwidth]{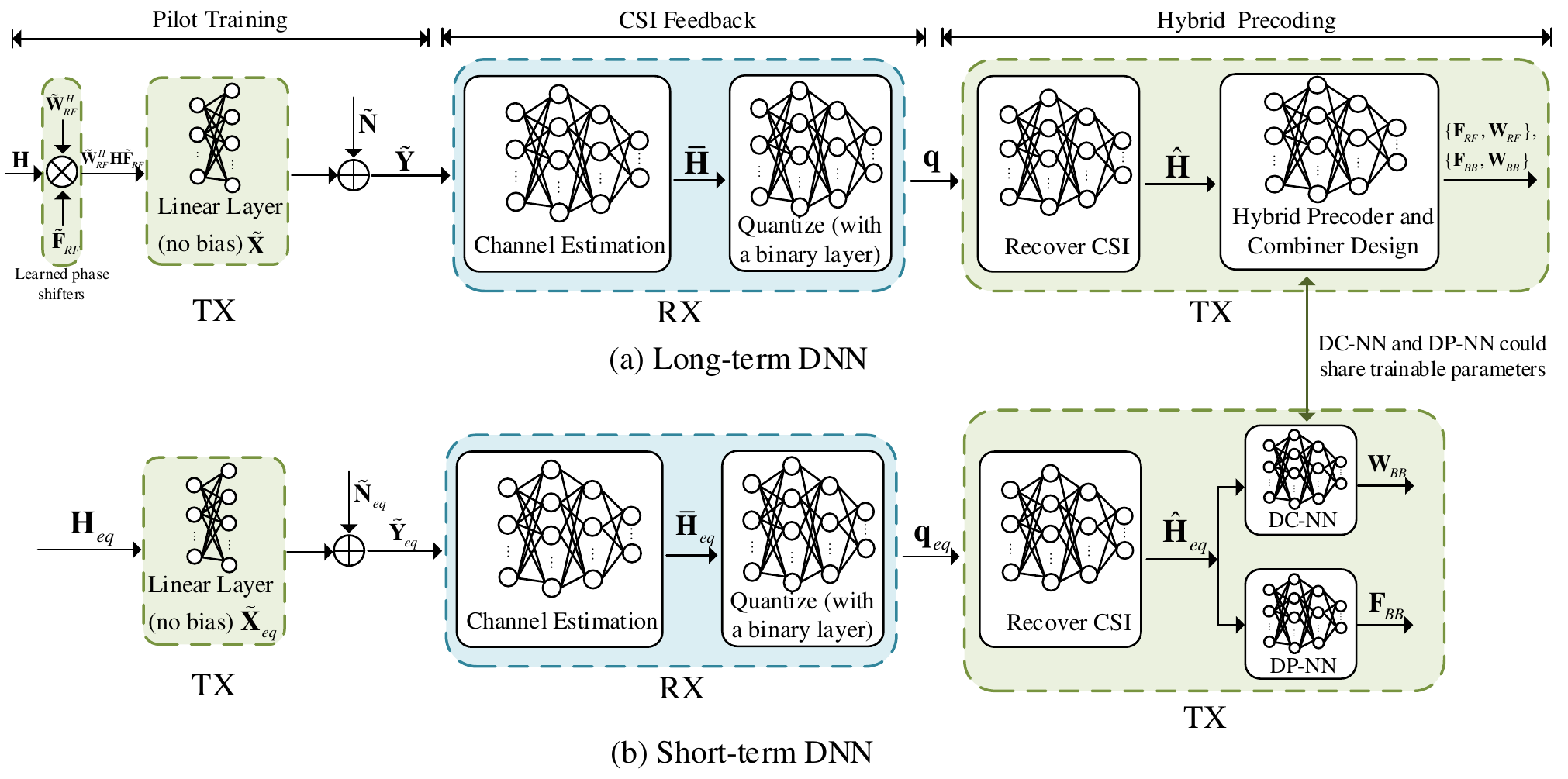}
\par\end{centering}
\caption{The architecture of the proposed two-timescale DNN that represents the end-to-end pilot training, CSI feedback, and hybrid precoding in an FDD mmWave MIMO system: (a) Long-term DNN that designs the hybrid precoder and combiner with full CSI samples; (b) Short-term DNN that designs the digital precoder and combiner with low-dimensional equivalent CSI matrices.}
\label{TwoTimeNN}
\end{figure*}

In this section, we propose a deep learning framework to achieve the joint design of the modules in Fig. \ref{FrameCSI}. Based on this proposed framework, a two-timescale DNN composed of a long-term DNN and a short-term DNN is developed to address problem \eqref{Problem}.

\subsection{Deep Learning Framework}
We aim at developing a deep learning framework for providing good BER performance with short pilot length $L$ and small number of feedback bits $B$.
The DNNs are employed to imitate an FDD mmWave MIMO system, which consists of the following stages: pilot training, CSI estimation and feedback, CSI recovery, hybrid precoding, and data transmission. 
We will present how to jointly design the training pilots $\{ \tilde{\mathbf{X}}, \tilde{\mathbf{W}}_{RF}, \tilde{\mathbf{F}}_{RF} \}$, CSI feedback scheme $\mathcal{F}(\cdot)$, and hybrid precoding scheme $\mathcal{G}(\cdot)$ in the long-term DNN designed in Section \ref{SecPilot}, \ref{SecFeed}, and \ref{Hybrid}, respectively. Fig. \ref{TwoTimeNN}(a) presents the block diagram of the deep learning framework designed for the aforementioned process, where the detailed architecture for hybrid precoder and combiner design and data transmission are presented in Fig. \ref{BlackboxPrecoder}. 
As seen in Fig. \ref{TwoTimeNN}(a), we employ a DNN at the RX whose inputs are the received pilots $\tilde{\mathbf{Y}}$ and outputs are feedback bits $\mathbf{q}$. At the TX, we train the pilots $\{ \tilde{\mathbf{X}}, \tilde{\mathbf{W}}_{RF}, \tilde{\mathbf{F}}_{RF} \}$ and apply a DNN whose inputs are feedback bits $\mathbf{q}$ and outputs are hybrid precoders and combiners $\{ \mathbf{F}_{RF}, \mathbf{F}_{BB}, \mathbf{W}_{RF}, \mathbf{W}_{BB} \}$.
Compared with the conventional scheme in Fig. \ref{FrameCSI}, we see that each module is replaced by a DNN. These DNNs can be jointly trained with the end-to-end bit-wise cross entropy (BCE) loss function, as illustrated in Section \ref{Hybrid}.
 
To further reduce the signaling overhead and CSI mismatch caused by the transmission delay, a two-timescale DNN composed of a long-term DNN and a short-term DNN is developed. In particular, the long-term DNN directly applies the architecture of the proposed deep learning framework in Fig. \ref{TwoTimeNN}(a) and it runs in the last time slot of each frame. In comparison, the short-term DNN in Fig. \ref{TwoTimeNN}(b) is modified based on the framework in Fig. \ref{TwoTimeNN}(a) and is implemented at each time slot. 
In particular, the short-term training pilots $\tilde{\mathbf{X}}_{eq}$, CSI feedback scheme $\mathcal{F}_{eq}(\cdot)$, and digital precoding scheme $\mathcal{G}_{eq}(\cdot)$ are jointly designed in the short-term DNN of Section \ref{SecPilot}, \ref{SecFeed}, and \ref{Hybrid}, respectively. 
As seen in Fig. \ref{TwoTimeNN}(b), we employ a DNN at the RX whose inputs are the received pilots $\tilde{\mathbf{Y}}_{eq}$ and outputs are the feedback bits $\mathbf{q}_{eq}$. At the TX, we train the pilots $\tilde{\mathbf{X}}_{eq}$ and apply a DNN whose inputs are feedback bits $\mathbf{q}_{eq}$ and outputs are digital precoders and combiners $\{ \mathbf{F}_{BB}, \mathbf{W}_{BB} \}$.
The communication process is summarized in Section \ref{SecProcess}.
In the following, we show the details of each module in Fig. \ref{TwoTimeNN}.

\subsection{Pilot Training} \label{SecPilot}
For pilot training, the RX needs to estimate the low-dimensional  equivalent CSI matrix $\mathbf{H}_{eq}$ in the first $T_{s}-1$ time slots of a frame and estimate the full CSI matrix $\mathbf{H}$ in the last time slot of this frame. 

\subsubsection{Pilot Training in the Long-Term DNN}
To estimate the full CSI matrix $\mathbf{H}$, the TX sends the training pilot matrix $\tilde{\mathbf{X}}\in \mathbb{C}^{N_{t}^{RF} \times L}$  modulated by the analog precoder $\tilde{\mathbf{F}}_{RF}\in \mathbb{C}^{N_t \times N^{RF}_{t}}$, where $L$ denotes the pilot length. Subsequently, the received pilot signal matrix processed by the analog combiner $\tilde{\mathbf{W}}_{RF}\in \mathbb{C}^{N_r \times N^{RF}_{r}}$ is expressed as 
\begin{equation}
\tilde{\mathbf{Y}} = \tilde{\mathbf{W}}_{RF}^{H} \mathbf{H} \tilde{\mathbf{F}}_{RF}\tilde{\mathbf{X}} + \tilde{\mathbf{N}}, 
\end{equation}
where $\tilde{\mathbf{N}}=\tilde{\mathbf{W}}_{RF}^{H}\mathbf{N}$, and $\mathbf{N}\in \mathbb{C}^{N_{r} \times L}$ denotes an AWGN matrix. 

To model the pilot training process and find the optimal pilots for estimation of $\mathbf{H}$, the input and output of this DNN are $\mathbf{H}$ and $\tilde{\mathbf{Y}}$, respectively, and the trainable parameters are $\{\tilde{\mathbf{X}}, \tilde{\mathbf{F}}_{RF}, \tilde{\mathbf{W}}_{RF}\}$. 
Compared with conventional approaches that apply a Gaussian pilot for $\tilde{\mathbf{X}}$ and select the columns from the DFT matrix for $\{\tilde{\mathbf{F}}_{RF}, \tilde{\mathbf{W}}_{RF}\}$ \cite{ChannelEsi00}, the trained $\{\tilde{\mathbf{X}}, \tilde{\mathbf{F}}_{RF}, \tilde{\mathbf{W}}_{RF}\}$ could achieve better channel estimation performance since they are trained adapt to the current CSI statistics.
To ensure that $\tilde{\mathbf{F}}_{RF}$ and $\tilde{\mathbf{W}}_{RF}$ satisfy the constant modulus constraints, we set the elements of these two matrices as trainable parameters which are divided by the absolute value, e.g., $\dfrac{ [\tilde{\mathbf{F}}_{RF}]_{ij} }{ | [\tilde{\mathbf{F}}_{RF}]_{ij} | }$. To guarantee that the pilot matrix $\tilde{\mathbf{X}}$ meets the transmit power constraint \eqref{Pilots}, we scale $\tilde{\mathbf{X}}$ such that $\|\tilde{\mathbf{x}}_{l} \|^{2}= P, \forall l$, where $\tilde{\mathbf{x}}_{l}$ (the $l$-th column of $\tilde{\mathbf{X}}$) denotes the transmitted pilots in the $l$-th pilot transmission.

Note that we can change $\tilde{\mathbf{F}}_{RF}$ and $\tilde{\mathbf{W}}_{RF}$ for $L$ pilots transmission and express the channel estimation process as
\begin{equation} 
\tilde{\mathbf{y}}_{l} = \tilde{\mathbf{W}}_{RF,l}^{H} \mathbf{H}\tilde{\mathbf{F}}_{RF,l}\tilde{\mathbf{x}}_{l} + \tilde{\mathbf{n}}_{l},
\end{equation}
where $\tilde{\mathbf{F}}_{RF,l}\in \mathbb{C}^{N_t \times N^{RF}_{t}}$ and $\tilde{\mathbf{W}}_{RF,l}\in \mathbb{C}^{N_r \times N^{RF}_{r}}, l=1, 2, \cdots, L,$ represent the analog precoder and combiner in the pilot training stage with the $l$-th pilot transmission, respectively. In addition, $\tilde{\mathbf{x}}_{l}$ and $\tilde{\mathbf{y}}_{l}$ denote the $l$-th column of the transmitted pilot matrix $\tilde{\mathbf{X}}$ and the received pilot matrix $\tilde{\mathbf{Y}}$, respectively. Here $\tilde{\mathbf{F}}_{RF,l}$ and $\tilde{\mathbf{W}}_{RF,l}$ are set as trainable parameters of the DNN. In this way, the RF precoder and combiner are different in each pilot transmission, which could excite several angular modes of the mmWave MIMO channel and achieve better system performance.

\subsubsection{Pilot Training in the Short-Term DNN}
To estimate the low-dimensional equivalent CSI matrix $\mathbf{H}_{eq}$, the TX sends the training pilot matrix $\tilde{\mathbf{X}}_{eq}\in \mathbb{C}^{N_{t}^{RF} \times L}$. The received pilot signal matrix at the RX is given by
\begin{equation}
\tilde{\mathbf{Y}}_{eq} = \mathbf{H}_{eq}\tilde{\mathbf{X}}_{eq} + \tilde{\mathbf{N}}_{eq}, 
\end{equation} 
where $\mathbf{H}_{eq} = \mathbf{W}_{RF}^{H} \mathbf{H}\mathbf{F}_{RF}$, $\tilde{\mathbf{N}}_{eq} = \mathbf{W}_{RF}^{H}\mathbf{N}$, and $\mathbf{N}\in \mathbb{C}^{N_{r} \times L}$ denotes an AWGN matrix. 

To model the pilot training process for the estimation of $\mathbf{H}_{eq}$, the input and output of this DNN are $\mathbf{H}_{eq}$ and $\tilde{\mathbf{Y}}_{eq}$, respectively, and its trainable parameter is $\tilde{\mathbf{X}}_{eq}$. The analog precoder $\mathbf{F}_{RF}$ and combiner $\mathbf{W}_{RF}$ in the short-term DNN are not trained but set as the values optimized at the hybrid precoding stage in the former frame, which will be further illustrated in Section \ref{Hybrid}. Hence, $\mathbf{F}_{RF}$ and $\mathbf{W}_{RF}$ are part of the input $\mathbf{H}_{eq}$. We scale $\tilde{\mathbf{X}}_{eq}$ to meet the transmit power constraint in the same way as that in the long-term DNN. 

In contrast to conventional channel estimation approaches, here the RX does not need to know the original pilot matrices $\{\tilde{\mathbf{X}}, \tilde{\mathbf{X}}_{eq}, \tilde{\mathbf{F}}_{RF}, \tilde{\mathbf{W}}_{RF}\}$ sent by the TX. This is because the pilot matrices are set as the trainable parameters of the DNN and are trained to be adapted to the current CSI statistics. The proposed DNN extracts the useful information from the received pilot matrices $\{ \tilde{\mathbf{Y}}, \tilde{\mathbf{Y}}_{eq} \}$ and the RX estimates the CSI matrices $\{\mathbf{H}, \mathbf{H}_{eq}\}$ only based on $\{ \tilde{\mathbf{Y}}, \tilde{\mathbf{Y}}_{eq} \}$ through the DNN. 

\subsection{CSI Feedback}   \label{SecFeed} 
The RX feeds back the quantized bits of the equivalent CSI matrix $\mathbf{H}_{eq}$ in the first $T_{s}-1$ time slots of a frame and those of the full CSI matrix $\mathbf{H}$ in the last time slot of this frame. 

\subsubsection{CSI Feedback in the Long-Term DNN}
The RX estimates the CSI matrix $\mathbf{H}$ based on the received pilot signal matrix $\tilde{\mathbf{Y}}$. Subsequently, the RX extracts the useful information and feeds back that information as $B$ bits to the TX for hybrid precoding. These two steps can be represented by a $R$-layer fully-connected (FC) DNN, where the feedback bits of the RX are given by
\begin{equation}
\mathbf{q} = \textrm{sgn} \big( \mathbf{W}_{R}\sigma_{R-1}\big( \cdots \sigma_{1} \big( \mathbf{W}_{1}\bar{\mathbf{y}}+\mathbf{b}_{1}  \big) \cdots \big) + \mathbf{b}_{R} \big). 
\end{equation}
Here $\mathbf{q}\in \{\pm 1\}^{B}$, $\tilde{\mathbf{y}}\triangleq \textrm{Vec}(\tilde{\mathbf{Y}})$ denotes the vectorization of matrix $\tilde{\mathbf{Y}}$, and the input of DNN is the real representation of $\tilde{\mathbf{y}}$, i.e., $\bar{\mathbf{y}} \triangleq [\Re(\tilde{\mathbf{y}}^{T}), \Im(\tilde{\mathbf{y}}^{T})]^{T}$. Note that $\{ \mathbf{W}_{r},\mathbf{b}_{r} \}_{r=1}^{R}$ denote the set of trainable parameters, $\sigma_{r}$ represents the activation function for the $r$-th layer, and the sign function $\textrm{sgn}(\cdot)$ is the activation function of the last layer (binary layer) to generate binary feedback bits for each element of $\mathbf{q}$. 

\subsubsection{CSI Feedback in the Short-Term DNN}
The feedback for $\mathbf{H}_{eq}$ follows the same procedure. The RX estimates $\mathbf{H}_{eq}$ based on the received pilot matrix $\tilde{\mathbf{Y}}_{eq}$ and extracts useful information for feedback with $B_{eq}$ bits. These two steps are represented by a $R_{eq}$-layer FC DNN and the feedback bits are given by
\begin{equation}
\mathbf{q}_{eq} = \textrm{sgn} \big( \mathbf{W}_{R_{eq}}\sigma_{R_{eq}-1}\big( \cdots \sigma_{1} \big( \mathbf{W}_{1}\bar{\mathbf{y}}_{eq}+\mathbf{b}_{1}  \big) \cdots \big) + \mathbf{b}_{R_{eq}} \big), 
\end{equation}
where $\tilde{\mathbf{y}}_{eq}\triangleq \textrm{Vec} (\tilde{\mathbf{Y}}_{eq})$ denotes the vectorization of matrix $\tilde{\mathbf{Y}}_{eq}$, and the input of DNN is the real representation of $\tilde{\mathbf{y}}_{eq}$, i.e., $\bar{\mathbf{y}}_{eq} \triangleq [\Re(\tilde{\mathbf{y}}^{T}_{eq}), \Im(\tilde{\mathbf{y}}^{T}_{eq})]^{T}$.
Note that the feedback bits $\mathbf{q}_{eq}\in \{\pm 1\}^{B_{eq}}$ have much lower dimension, i.e., $B_{eq}< B$, since the dimension of $\mathbf{H}_{eq}$ is much lower than that of $\mathbf{H}$. Thus, we can employ a DNN with a smaller number of layers $R_{eq}$ and low-dimensional parameters, i.e., $\{ \mathbf{W}_{r},\mathbf{b}_{r} \}_{r=1}^{R_{eq}}$, to obtain the feedback bits $\mathbf{q}_{eq}$. 

\subsection{Hybrid Precoder and Combiner Design} \label{Hybrid}

\begin{figure*}[t]
\begin{centering}
\includegraphics[width=0.8\textwidth]{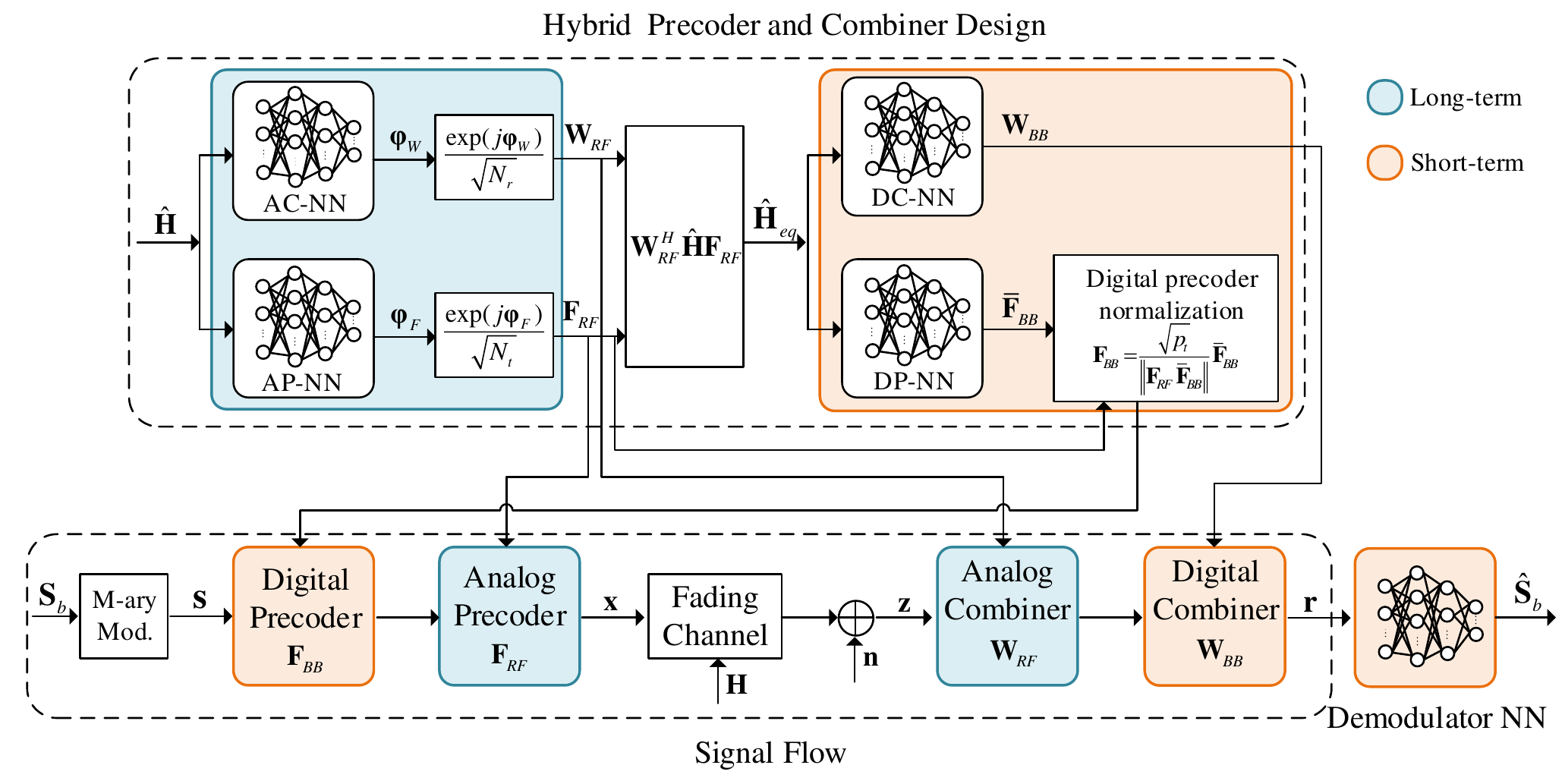}
\par\end{centering}
\caption{The DNN architecture for hybrid precoder and combiner design in the data transmission stage.}
\label{BlackboxPrecoder}
\end{figure*}

In the first $T_{s}-1$ time slots of a frame, we employ the short-term DNN to update the digital precoder and combiner $\{\mathbf{F}_{BB}, \mathbf{W}_{BB}\}$ based on $\mathbf{q}_{eq}$. In the last time slot of the frame, we apply the long-term DNN to update the hybrid precoder and combiner $\{\mathbf{F}_{RF}, \mathbf{F}_{BB}, \mathbf{W}_{RF}, \mathbf{W}_{BB}\}$ based on $\mathbf{q}$. We assume that the transmission of feedback bits $\mathbf{q}_{eq}$ and $\mathbf{q}$ between the RX and TX are error-free. 
 
\subsubsection{Hybrid Precoder and Combiner Design in the Long-Term DNN}
At the last time slot of each frame, the TX collects the feedback bits $\mathbf{q}$ to recover the full CSI matrix. Then, the TX designs the hybrid precoder and combiner based on the recovered CSI matrix $\hat{\mathbf{H}}$ with a DNN. 
Inspired from the single-timescale deep learning-based hybrid precoding with perfect CSI proposed in \cite{HybridDNN}, we design the hybrid precoding in a two-timescale manner.
As shown in Fig. \ref{BlackboxPrecoder}, it includes five FC sub-NNs, i.e., the analog precoder NN (AP-NN), digital precoder NN (DP-NN), analog combiner NN (AC-NN), and digital combiner NN (DC-NN), as well as a demodulator NN. Specifically, $\hat{\mathbf{H}}$ is firstly converted into a $2 N_{t} N_{r} \times 1$ real-valued vector and then input into the AP-NN and AC-NN to generate $\bm{\varphi}_{F} \in \mathbb{R}^{N_{t}N^{RF}_{t}\times 1 }$ and $\bm{\varphi}_{W}\in \mathbb{R}^{N_{r}N^{RF}_{r}\times 1 }$ for phase shifters in the TX and RX, respectively. Two complex-valued vectors with constant modulus elements are then obtained as
\begin{equation} \label{PhaseShifter}
\bar{\mathbf{f}}_{RF}=\frac{1}{\sqrt{N_{t}}}e^{j\bm{\varphi}_{F}},\quad
\bar{\mathbf{w}}_{RF}=\frac{1}{\sqrt{N_{r}}}e^{j\bm{\varphi}_{W}}.
\end{equation}
Then, $\mathbf{F}_{RF}$ and $\mathbf{W}_{RF}$ can be written as
\begin{equation} \label{FRF}
\mathbf{F}_{RF}=\mathcal{J}_{v\rightarrow m}(\bar{\mathbf{f}}_{RF}), \quad
\mathbf{W}_{RF}=\mathcal{J}_{v\rightarrow m}(\bar{\mathbf{w}}_{RF}),
\end{equation}
where $\mathcal{J}_{v\rightarrow m}$ represents the operation that reshapes a vector into a matrix. The resulting $\mathbf{F}_{RF}$ and $\mathbf{W}_{RF}$ along with $\hat{\mathbf{H}}$ are employed to generate a low-dimensional equivalent CSI as
\begin{equation} \label{Heq}
\hat{\mathbf{H}}_{eq}=\mathbf{W}_{RF}^{H}\hat{\mathbf{H}}\mathbf{F}_{RF}. 
\end{equation}

Next, $\hat{\mathbf{H}}_{eq}\in \mathbb{C}^{N^{RF}_{r}\times N^{RF}_{t}}$ is converted to a $2N^{RF}_{t}N^{RF}_{r}\times 1$ real-valued vector and inputted into the DP-NN and DC-NN, the outputs of which are $\{ \bar{\mathbf{w}}_{BB,re},\bar{\mathbf{w}}_{BB,im} \}$ and $\{ \bar{\mathbf{f}}_{BB,re},\bar{\mathbf{f}}_{BB,im} \}$, respectively. Finally, $\mathbf{W}_{BB}$ and $\mathbf{F}_{BB}$ are computed as
\begin{equation} \label{WBB}
\begin{aligned}
& \mathbf{W}_{BB}=\mathcal{J}_{v\rightarrow m}(\bar{\mathbf{w}}_{BB,re}+j\bar{\mathbf{w}}_{BB,im}), \\
&\bar{\mathbf{F}}_{BB}=\mathcal{J}_{v\rightarrow m}(\bar{\mathbf{f}}_{BB,re}+j\bar{\mathbf{f}}_{BB,im}).
\end{aligned}
\end{equation}
The final digital precoder that meets the power constraint \eqref{Power} follows using $\bar{\textbf{F}}_{BB}$ in \eqref{WBB} and $\textbf{F}_{RF}$ in \eqref{FRF}:
\begin{equation} \label{Normalization}
\mathbf{F}_{BB} = \dfrac{\sqrt{P_{T}}}{\| \mathbf{F}_{RF} \bar{\mathbf{F}}_{BB}\|} \bar{\mathbf{F}}_{BB}.
\end{equation}

\subsubsection{Digital Precoder and Combiner Design in the Short-Term DNN}
In the first $T_{s}-1$ time slots of the frame, the TX collects the feedback bits $\mathbf{q}_{eq}$ to recover the low-dimensional equivalent CSI $\hat{\mathbf{H}}_{eq}$. Then, the TX designs the digital precoder $\mathbf{F}_{BB}$ and combiner $\mathbf{W}_{BB}$ based on the recovered equivalent CSI matrix $\hat{\mathbf{H}}_{eq}$ with a DNN, while the analog precoder $\mathbf{F}_{RF}$ and combiner $\mathbf{W}_{RF}$ are fixed. As presented in Fig. \ref{TwoTimeNN}(b), the short-term DNN consisting of a DP-NN and a DC-NN generates the digital precoder and combiner, respectively. Finally, $\mathbf{F}_{BB}$ and $\mathbf{W}_{BB}$ are obtained based on \eqref{WBB}-\eqref{Normalization}.

\subsubsection{Signal Flow}
The goal of offline training is to learn the trainable parameters $\bm{\Theta}$ of the DNNs based on the training samples with the input tuple $\{ \mathbf{H}, \mathbf{n}, \mathbf{S}_{b} \}$, and the label $\mathbf{S}_{b}$. We assume certain distributions of the CSI and noise and accordingly generate a large number of CSI and noise realizations for training. 

The signal flow in Fig. \ref{BlackboxPrecoder} simulates the process from the transmitted signal $\mathbf{S}_{b}$ to the recovered signal  $\hat{\mathbf{S}}_{b}$, over the wireless fading channel $\mathbf{H}$, with an AWGN vector $\mathbf{n}$, where the hybrid precoder and combiner $\{\mathbf{F}_{RF}, \mathbf{F}_{BB}, \mathbf{W}_{RF}, \mathbf{W}_{BB}\}$ are generated following the steps in \eqref{PhaseShifter}-\eqref{Normalization}.
The signal model \eqref{SignalModel} is executed through the DNNs by using the input tuple $\{ \mathbf{H}, \mathbf{n}, \mathbf{S}_{b} \}$ and the hybrid precoder and combiner to yield the received signal $\mathbf{r}$. By augmenting the real and imaginary parts, $\mathbf{r}$ is converted to a real-valued vector and input into the demodulator NN to produce the recovered signal $\hat{\mathbf{S}}_{b}$. 
By minimizing the end-to-end BCE between $\mathbf{S}_{b}$ and $\hat{\mathbf{S}}_{b}$, the trainable parameters $\bm{\Theta}$ of the DNNs are updated iteratively by the stochastic gradient descent (SGD).
In the deployment and testing stage, the modules in the signal flow are replaced by the hybrid precoder and combiner optimized by the DNNs. 

\subsubsection{BCE Loss Function and BER}
The BCE shown below is applied as the loss function,
\begin{equation} \label{BCE}
\begin{aligned}
\mathcal{L}(\bm{\Theta}) = - \frac{1}{|\mathcal{B}|} \sum\limits_{\mathbf{S}_{b}\in \mathcal{B}} \sum\limits_{i=1}^{N_{s}} \sum\limits_{j=1}^{\log_{2}M}\bigg( [\mathbf{S}_{b}]_{i,j} \ln( [\hat{\mathbf{S}}_{b}(\bm{\Theta})]_{i,j} ) \\
+ (1-[\mathbf{S}_{b}]_{i,j})\ln(1-[\hat{\mathbf{S}}_{b}(\bm{\Theta})]_{i,j} ) \bigg),
\end{aligned}
\end{equation}
where $\mathcal{B}$ denotes the training symbol dataset and $\mathbf{S}_{b}$ is a transmitted symbol matrix consisting of the binary bits with dimension $N_{s}\times \log_{2}M$. The $[\hat{\mathbf{S}}_{b}(\bm{\Theta})]_{i,j} \in [0,1]$ denotes the recovered symbol matrix, which indicates the probability of the transmitted bit to be $1$ and is expressed as the function of the parameter set $\bm{\Theta}$ of all the DNNs. Note that maximizing the BCE essentially maximizes an achievable rate that we can obtain with a standard bit-metric decoder \cite{DNNSystem}.

Recalling the optimization problem in \eqref{Problem}, the BER over the training dataset can be defined as
\begin{equation} \label{BER}
\begin{aligned}
P_{e}(\bm{\Theta}) & \triangleq P_{e}(\mathbf{F}_{RF}, \mathbf{F}_{BB}, \mathbf{W}_{RF}, \mathbf{W}_{BB})   \\
&= \dfrac{ \sum\limits_{\mathbf{S}_{b}\in \mathcal{B}}  \sum\limits_{i=1}^{N_{s}} \sum\limits_{j=1}^{\log_{2}M} \big| [\mathbf{S}_{b}]_{i,j} - [\hat{\mathbf{S}}_{b,1hot}(\bm{\Theta})]_{i,j} \big| } { |\mathcal{B}|N_{s}\log_{2}M },
\end{aligned}
\end{equation}
where $[\hat{\mathbf{S}}_{b,1hot}(\bm{\Theta})]_{i,j}=0$ for $[\hat{\mathbf{S}}_{b}(\bm{\Theta})]_{i,j}<0.5$ and $[\hat{\mathbf{S}}_{b,1hot}(\bm{\Theta})]_{i,j}=1$ otherwise.

\subsubsection{The Deployment of DNNs}
The proposed hybrid precoding DNNs consist of the AP-NN, DP-NN, AC-NN, and DC-NN, which should be jointly trained. Then, there are two kinds of deployment methods for the trained hybrid precoding DNNs: (i) the four NNs are all deployed at the TX; and (ii) the AP-NN and DP-NN are deployed at the TX, while the AC-NN and DC-NN are deployed at the RX. As for the first method, the TX needs to design the precoders and combiners through these four trained DNNs. Then, the TX needs to feed forward the designed combiners to the RX. As for the second method, the precoders are designed at the TX through the AP-NN and DP-NN, and the combiners are designed at the RX via the AC-NN and DC-NN. The first approach does not require the RX to have efficient computing resources, while the second one does not require the TX to feed forward the designed combiners to the RX.

\subsection{Two-timescale Communication Process} \label{SecProcess}

\begin{figure}[t]
\begin{centering}
\includegraphics[width=0.5\textwidth]{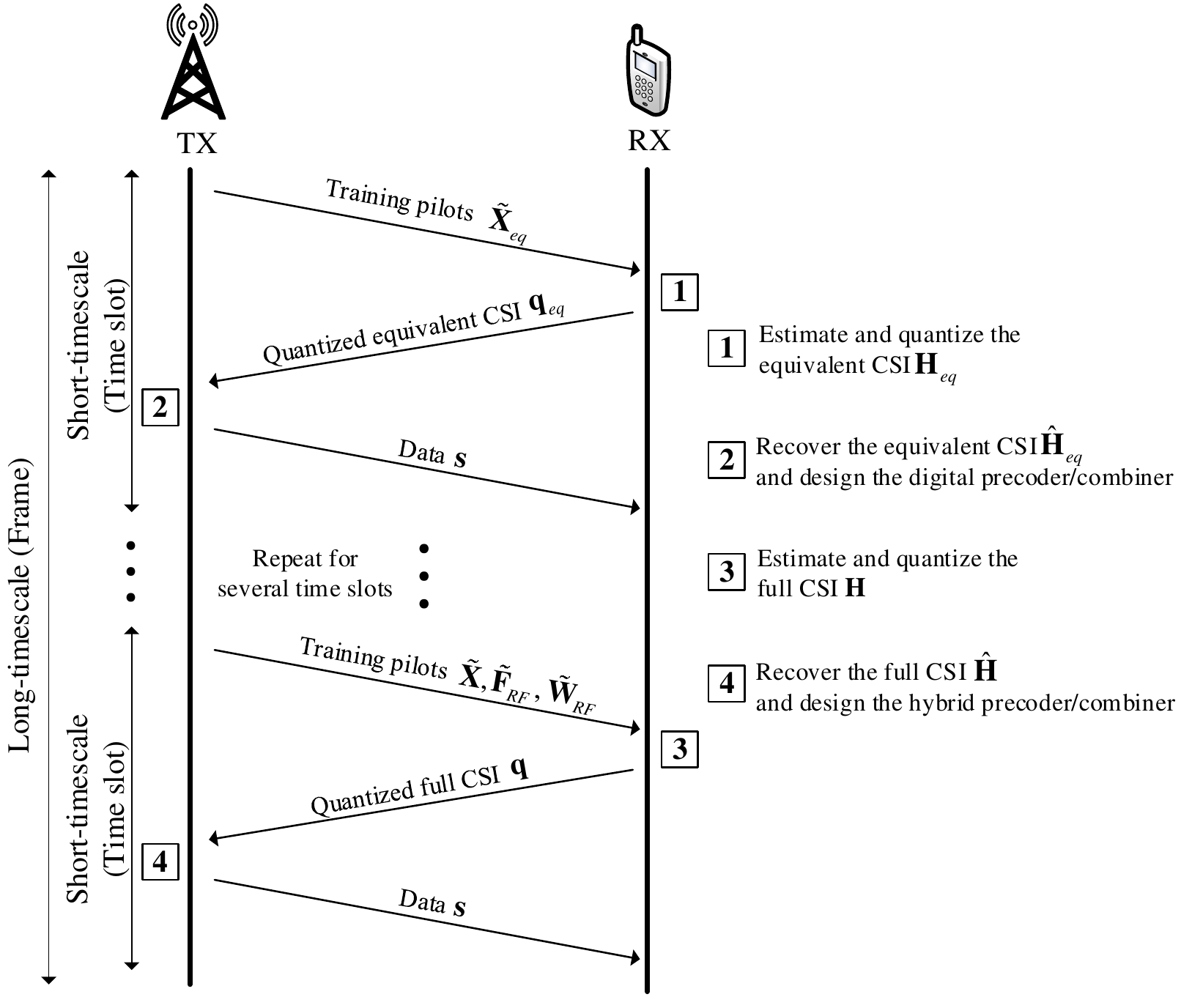}
\par\end{centering}
\caption{The communication process of the FDD end-to-end mmWave MIMO system.}
\label{FlowChart}
\end{figure}

Fig. \ref{FlowChart} shows the communication process of the end-to-end FDD mmWave MIMO system with two-timescale hybrid precoding. In the first $T_{s}-1$ time slots of a frame, the TX transmits the training pilots $\tilde{\mathbf{X}}_{eq}$ to the RX. Then, the RX estimates and quantizes the equivalent CSI matrix $\mathbf{H}_{eq}$ and feeds the quantized bits $\mathbf{q}_{eq}$ back to the TX. Subsequently, the TX recovers $\hat{\mathbf{H}}_{eq}$ and designs the digital precoder and combiner $\{\mathbf{F}_{BB}, \mathbf{W}_{BB}\}$ while keeping the analog precoder and combiner $\{\mathbf{F}_{RF}, \mathbf{W}_{RF}\}$ unchanged. Finally, the data is transmitted by following the signal flow as shown in Fig. \ref{BlackboxPrecoder}. 
In comparison, in the last time slot of the frame, the TX first transmits the training pilots $\{\tilde{\mathbf{X}}, \tilde{\mathbf{F}}_{RF}, \tilde{\mathbf{W}}_{RF}\}$ to the RX which then estimates and quantizes the full CSI matrix $\mathbf{H}$ and feeds the quantized bits $\mathbf{q}$ back to the TX. Then, the TX recovers $\hat{\mathbf{H}}$ and designs the hybrid precoder and combiner $\{\mathbf{F}_{BB}, \mathbf{F}_{RF}, \mathbf{W}_{BB}, \mathbf{W}_{RF}\}$. Finally, the actual data $\mathbf{s}$ is transmitted. The dimension of $\mathbf{q}_{eq}$ is much smaller than that of $\mathbf{q}$ due to the significantly reduced dimension of $\mathbf{H}_{eq}$ compared to $\mathbf{H}$.

The frame structure can be summarized as follows. A frame contains several time slots, and the structure of a time slot is composed of four parts: indicating bits, pilot symbols, feedback bits, and transmission data. In particular, the indicator bits indicate: (i) whether the current time slot employs the long-term DNN or a short-term DNN, (ii) whether the current CSI statistics change, and (iii) whether the CSI statistics change faster or slower. When the CSI statistics change, since it will not change significantly within a short time, the CSI samples from the changed CSI statistics are collected for fine-tuning (online training and transfer learning) based on the previously trained DNN, which converges fast within several time slots. If the CSI statistics change faster or slower, the frame and time slot length needs to be adjusted adaptively. When the CSI statistics change faster, the length of frame and time slot needs to be shortened to obtain more high-dimensional original CSI samples to track the change. 

We note in conclusion that the proposed two-timescale DNN in FDD mode could be flexibly extended to time-division duplex (TDD) mode by: (i) removing the CSI feedback part; and (ii) modifying the pilot training stage by letting the RX send the pilots. Since the uplink channel and downlink channel follow reciprocity in TDD mode, to acquire the downlink CSI matrix for designing the hybrid precoders in the downlink data transmission stage, the TX could first estimate the uplink CSI matrix based on the received pilots sent by the RX. Then, the downlink CSI matrix can be obtained at the TX based on channel reciprocity. Thus, CSI feedback is not required in TDD mode. 

\section{Practical Implementation} \label{Impleme}
In this section, we describe the architecture and training method for the proposed two-timescale DNN with a binary layer for practical implementation. 

\subsection{The Architecture of the Proposed DNN}
Generally, we employ the FC DNN with a non-linear function ``Sigmoid" in the last layer and the ``ReLU" in the other layers. Specifically, a $4$-layer DNN is applied for channel estimation and the number of neurons in different layers are $[l_{1}, l_{2}, l_{3}, l_{4}]=[N_{r}L, 256, 128, N_{r}N_{t}]$. As for the quantization and CSI recovery, a $3$-layer DNN with $[l_{1}, l_{2}, l_{3}]=[N_{r}N_{t}, 128, B]$ and a $4$-layer DNN with $[l_{1}, l_{2}, l_{3}, l_{4}]=[B, 256, 128, N_{r}N_{t}]$ are employed, respectively. For the analog precoder and combiner, we employ the $4$-layer DNN with $[l_{1}, l_{2}, l_{3}, l_{4}]=[N_{r}N_{t}, 256, 128, N_{r}N_{r}^{RF}]$ for the AC-NN and $[l_{1}, l_{2}, l_{3}, l_{4}]=[N_{r}N_{t}, 256, 128, N_{t}N_{t}^{RF}]$ for the AP-NN. Since the digital precoder and combiner have much smaller dimensions than those of the analog ones, we apply the $4$-layer DNN with much reduced number of neurons, i.e., $[l_{1}, l_{2}, l_{3}, l_{4}]=[N_{r}^{RF}N_{t}^{RF}, 64, 32, N_{r}^{RF}N_{s}]$ for the DC-NN and $[l_{1}, l_{2}, l_{3}, l_{4}]=[N_{r}^{RF}N_{t}^{RF}, 64, 32, N_{t}^{RF}N_{s}]$ for the DP-NN. In addition, a $4$-layer DNN is designed for the NN demodulator with $[l_{1}, l_{2}, l_{3}, l_{4}]=[2N_{s}, 64, 32, N_{s}\log_{2}M]$. Furthermore, we apply the batch normalization and the residual block in ``ResNet" to solve the problem of gradient vanishing and explosion, which improves the system performance.

\subsection{Implementation for Analog Precoding}

\begin{figure}[t]
\begin{centering}
\includegraphics[width=0.5\textwidth]{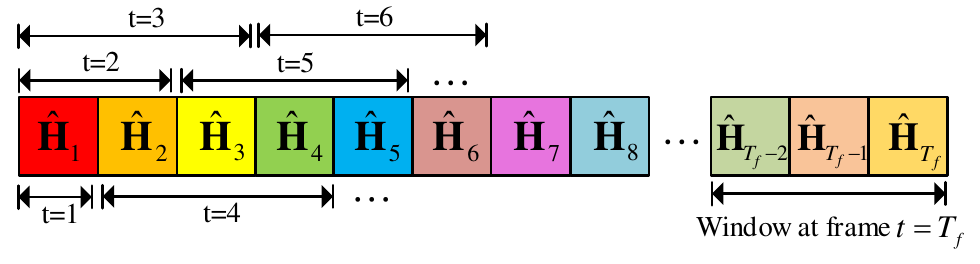}
\par\end{centering}
\caption{The sliding window of the full CSI with size $D=3$.}
\label{Window}
\end{figure}

Note that the long-term variables, i.e., the phases of analog precoder $\mathbf{F}_{RF}$ and combiner $\mathbf{W}_{RF}$ defined in \eqref{PhaseShifter}, should be adapted to the CSI statistics. Thus, they are optimized based on a sufficient number of full CSI samples $\mathbf{H}$. However, only one sample is obtained at each frame. 
Based on \cite{SSCA}, the long-term variables are updated by following moving average approach to take full advantage of these samples as
\begin{equation} \label{UpdateLongterm}
\bm{\varphi}_{F}^{t+1} = (1-\gamma_{t}) \bm{\varphi}_{F}^{t} + \gamma_{t} \bar{\bm{\varphi}}_{F}^{t}, \quad 
\bm{\varphi}_{W}^{t+1} = (1-\gamma_{t}) \bm{\varphi}_{W}^{t} + \gamma_{t} \bar{\bm{\varphi}}_{W}^{t}.
\end{equation}
Here $\bm{\varphi}_{F}^{t}$ and $\bar{\bm{\varphi}}_{F}^{t}$ denote the current phase of the analog precoder and the output of the AP-NN at the $t$-th frame, respectively, $\bm{\varphi}_{W}^{t}$ and $\bar{\bm{\varphi}}_{W}^{t}$ denote the current phase of the analog combiner and the output of the AC-NN at the $t$-th frame, respectively, and $\{\gamma_{t}, t=1, 2, \cdots \!, T_{f}\}$ denote a sequence of parameters selected to meet the conditions \cite{SSCA}: $\lim\limits_{t\rightarrow \infty} \gamma_{t} = 0$, $\sum_{t} \gamma_{t} = \infty$, and $\sum_{t} (\gamma_{t})^{2} < \infty$. 

Furthermore, to make the long-term variables better fit the CSI statistics and make full use of the full CSI samples, we employ a sliding window (buffer) $\mathcal{D}$ with size $D$ to store the previously recovered full CSI sample $\hat{\mathbf{H}}$ at each frame, as shown in Fig. \ref{Window}. The input of the AC-NN and AP-NN at the $t$-th frame is the matrix consisting of $D$ recovered full CSI samples from the frame $t-D+1$ to the current frame $t$, i.e., $\big[ \hat{\mathbf{H}}_{t-D+1}, \hat{\mathbf{H}}_{t-D+2}, \cdots, \hat{\mathbf{H}}_{t} \big]$.

\subsection{Training Method}
\subsubsection{DNN Training with a Binary Layer for CSI Feedback}
Since the derivative of the output of a binary neuron (the activation function is $\textrm{sgn}(\cdot)$) is $0$ almost everywhere, except the origin where the function is non-differentiable, the conventional back-propagation method cannot be directly applied to train the layers preceding the binary layer. A so-called straight-through (ST) estimator \cite{Lecture} has been proposed to address this issue, where the activation function of the binary layer is approximated by a smooth differentiable function in back-propagation. A variant of the ST estimator, referred to as sigmoid-adjusted ST, replaces the activation function $\textrm{sgn}(x)$ with $2\textrm{sigm}(x)-1$, where $\textrm{sigm}(x)=1/(1+\textrm{exp}(-x))$ denotes the sigmoid function. The performance of the sigmoid-adjusted ST estimator can be further improved by employing slope-annealing \cite{STestimator}, where the slope of the sigmoid function is gradually increased as the training progress. Particularly, the sigmoid-adjusted ST with slope-annealing estimator approximates the sign function with a scaled sigmoid function as
\begin{equation}  \label{SigmoidST}
2\textrm{sigm}(\alpha^{(i)}x)-1= \dfrac{2}{(1+\textrm{exp}(-\alpha^{(i)}x))}-1,
\end{equation}
where $\alpha^{(i)}$ denotes the annealing factor in the $i$-th epoch that satisfies $\alpha^{(i)}\geq \alpha^{(i-1)}$. 

\subsubsection{Training and Testing Procedures for the Two-Timescale DNN}
The DNN is trained in the two-timescale manner, where the short-term and long-term DNNs are trained alternately. Specifically, the short-term DNN is trained in the first $T_{s}-1$ time slots of a frame, whose inputs are the training samples $\{ \mathbf{H}, \mathbf{F}_{RF}, \mathbf{W}_{RF}, \mathbf{n}, \mathbf{S}_{b} \}$. Note that $\mathbf{F}_{RF}$ and $\mathbf{W}_{RF}$ corresponding to $\mathbf{H}$ are computed by the long-term DNN. In comparison, the long-term DNN is trained in the last time slot of the frame, where a batch of training samples $\{ \mathbf{H}, \mathbf{n}, \mathbf{S}_{b} \}$ are input into the long-term DNN and then it is trained by the SGD. 
Furthermore, the testing stage is executed in a similar way. In the first $T_{s}-1$ time slots of a frame, we input the pilot $\tilde{\mathbf{X}}_{eq}$ and perform the forward-propagation in the short-term DNN to compute $\{\mathbf{F}_{BB}, \mathbf{W}_{BB}\}$ based on \eqref{WBB}-\eqref{Normalization}. At the last time slot of each frame, we input the pilot $\tilde{\mathbf{X}}$ and perform the forward-propagation in the long-term DNN to compute $\{\mathbf{F}_{RF}, \mathbf{F}_{BB}, \mathbf{W}_{RF}, \mathbf{W}_{BB}\}$ according to \eqref{PhaseShifter}-\eqref{Normalization} and \eqref{UpdateLongterm}.

\section{Analysis of the Proposed DNN} \label{Analysis}
In this section, we develop a method for enhancing the generalization ability of the proposed DNN and analyze the signaling overhead of the proposed method by comparing it with existing schemes.

\subsection{Generalization Ability} \label{GeneralB}
The generalization ability of DNNs can be divided into two categories. The first category consists of parameters that only change the input distribution, e.g., the channel parameter $N_{cl}$, SNR, and noise statistics $\sigma_{n}^{2}$. For these parameters, the generalization ability can be enhanced by training under a variety of system parameters. By taking the SNR as an example, we train the proposed DNN over different values of SNR. The second category changes the input/output dimensions of the layers in the DNN, e.g., the number of feedback bits $B$, the length of training pilots $L$, and the number of antennas $(N_{t},N_{r})$ and RF chains $(N_{t}^{RF},N_{r}^{RF})$. Training a DNN to use for different system dimensions becomes much more challenging. In the following, we illustrate how to improve the generalization ability of the proposed DNN in the second category.

\subsubsection{Generalization to a Varying Number of Feedback Bits $B$} 
We aim at training a general DNN that can operate over a wide range of $B$ in practical systems \cite{FeedbackFDD}, where a training method with two steps is developed. First, we train a modified version of the proposed DNN, where the outputs of the CSI feedback DNN, i.e., $\mathbf{q}$, are not binary but real-valued within the range from $-1$ to $1$, generated by $P$ neurons with ``tanh" activation functions. The modified DNN is employed to obtain the pilot sequences and the channel estimation scheme. After the training of this modified DNN, we acquire the empirical probability distribution function (PDF) of the output of the ``tanh" layer, and then design an optimal scalar quantizer based on the Lloyd-Max algorithm for different values of quantization bits $Q$. 
In the second step, the DNN parameters at the RX are fixed while training the DNN parameters for the hybrid precoding at the TX. We apply different quantization resolutions to these $P$ signals to account for different feedback capacities $B$. Specifically, the TX receives a $Q$-bit quantized version of the $P$ signals from the RX, and the DNN at the TX aims at mapping these quantized signals to the hybrid precoding matrices. Note that the amount of feedback bits are $B=P\times Q$, thus by changing the different quantization levels $Q$, we can employ a trained DNN to operate for systems with different values of $B$.

\subsubsection{Generalization to a Varying Number of Pilot Length $L$}
The trained DNN with a larger value of $L_{0}$ can be directly employed to test the samples with a smaller value of $L_{1}$. Let us take the short-term DNN as an example, where the last $L_{0}-L_{1}$ columns of the received signal matrix $\tilde{\mathbf{Y}}_{eq}\in \mathbb{C}^{N_{r}^{RF} \times L_{0}}$ are set to be $\bm{0}$, or equivalently, the last $L_{0}-L_{1}$ columns of the training pilot matrix $\tilde{\mathbf{X}}_{eq}\in \mathbb{C}^{N_{t}^{RF} \times L_{0}}$ and noise matrix $\tilde{\mathbf{N}}_{eq}\in \mathbb{C}^{N_{r}^{RF} \times L_{0}}$ are set to be $\bm{0}$:
\begin{equation}      
\left[             
\begin{array}{c}   
\tilde{\mathbf{Y}}_{eq}^{'}, \mathbf{0}  \\ 
\end{array} \right] 
= \mathbf{H}_{eq}   
\left[              
\begin{array}{c}  
\tilde{\mathbf{X}}_{eq}^{'}, \mathbf{0}   \\ 
\end{array} \right]    
+ \left[  \begin{array}{c}   
\tilde{\mathbf{N}}_{eq}^{'}, \mathbf{0}  \\ 
\end{array} \right],    
\end{equation}
where $\tilde{\mathbf{X}}_{eq}^{'}\in \mathbb{C}^{N_{t}^{RF} \times L_{1}}$, $\tilde{\mathbf{Y}}_{eq}^{'}\in \mathbb{C}^{N_{r}^{RF} \times L_{1}}$, and $\tilde{\mathbf{N}}_{eq}^{'}\in \mathbb{C}^{N_{r}^{RF} \times L_{1}}$. Note that the generalization ability for $L$ in the long-term DNN can be analyzed similarly.

\subsubsection{Generalization to a Varying Number of $N_{t}^{RF}$ and $N_{r}^{RF}$}
The trained DNN with the system configuration $(N_{t_{0}}^{RF}, N_{r_{0}}^{RF})$ can be straightforwardly employed to test the samples with smaller values of $(N_{t_{1}}^{RF}, N_{r_{1}}^{RF})$, rather than training a new DNN. 
In the pilot training stage, to ensure that the input of the smaller system, i.e., $\mathbf{H}_{eq}^{'}\in \mathbb{C}^{N_{r_{1}}^{RF}\times N_{t_{1}}^{RF}}$, has the same dimension with that of the larger system, i.e., $\mathbf{H}_{eq}\in \mathbb{C}^{N_{r_{0}}^{RF}\times N_{t_{0}}^{RF}}$, we perform zero padding that adds $N_{t_{0}}^{RF}-N_{t_{1}}^{RF}$ zero columns and $N_{r_{0}}^{RF}-N_{r_{1}}^{RF}$ zero rows to $\mathbf{H}_{eq}^{'}$ as
\begin{equation}      
\left[             
\begin{array}{c}   
\tilde{\mathbf{Y}}_{eq}^{'}  \\  
\bm{0}  \\ 
\end{array} \right] 
= \left[              
\begin{array}{cc}  
\mathbf{H}_{eq}^{'} & \bm{0} \\  
\bm{0} & \bm{0} \\ 
\end{array} \right]   
\left[              
\begin{array}{c}  
\tilde{\mathbf{X}}_{eq}^{'}  \\  
\bm{0}  \\ 
\end{array} \right]    
+ \left[  \begin{array}{c}   
\tilde{\mathbf{N}}_{eq}^{'}  \\  
\bm{0}  \\ 
\end{array} \right],    
\end{equation}
where $\tilde{\mathbf{Y}}_{eq}^{'}\in \mathbb{C}^{N_{r_{1}}^{RF} \times L}$ denotes the received signal, $\tilde{\mathbf{N}}_{eq}^{'}\in \mathbb{C}^{N_{r_{1}}^{RF} \times L}$ is the noise matrix, and $\tilde{\mathbf{X}}_{eq}^{'}\in \mathbb{C}^{N_{t_{1}}^{RF} \times L}$ denotes the trained pilot matrix from the larger system $(N_{t_{0}}^{RF}, N_{r_{0}}^{RF})$. Note that only the first $N_{t_{1}}^{RF}$ rows of the trained pilot matrix, i.e., $\tilde{\mathbf{X}}_{eq}^{'}$, are employed for the smaller system $(N_{t_{1}}^{RF}, N_{r_{1}}^{RF})$.
 
Furthermore, we perform zero padding for the hybrid precoder and combiner. As for the analog precoding matrix $\mathbf{F}_{RF}\in \mathbb{C}^{N_t \times N^{RF}_{t}}$, we only need to set its last $N^{RF}_{t_{0}}-N^{RF}_{t_{1}}$ columns to be $\mathbf{0}$, i.e., $[\mathbf{F}_{RF}^{'},\mathbf{0}]$, where $\mathbf{F}_{RF}^{'}$ denotes the first $N_{t_{1}}^{RF}$ columns of the analog precoding matrix produced by the AP-NN of the larger system $(N_{t_{0}}^{RF}, N_{r_{0}}^{RF})$. The other precoders and combiners can be tackled in the same way. Moreover, the generalization ability of the long-term DNN and that regarding the system parameters $N_{t}$ and $N_{r}$ can be analyzed similarly.

\subsection{Analysis of the Signaling Overhead}
In this part, we analyze the signaling overhead of the proposed DNN in comparison with existing schemes. Consider a suprerframe consisting of $T_f$ frames, each of which contains $T_s$ time slots. The results are summarized as follows.

\begin{itemize}
\item Conventional single timescale approach: 
Note that $B_{c}$ denotes the number of quantization bits for each element of the CSI matrix $\mathbf{H}\in \mathbb{C}^{N_{r}\times N_{t}}$, hence the number of signaling bits in a superframe is given as $Q_{cs} = T_{f}T_{s}B_{c}N_{r}N_{t}$.
\item Conventional two-timescale approach: 
The RX feeds back the quantized bits of the equivalent CSI matrix $\mathbf{H}_{eq}\in \mathbb{C}^{N_{r}^{RF}\times N_{t}^{RF}}$ in the first $T_{s}-1$ time slots of a frame and feeds back those of the full CSI matrix $\mathbf{H}$ in the last time slot of each frame. Thus, the number of signaling bits of the two-timescale approach within a superframe is given by $Q_{ct} = T_{f}B_{c}\big( (T_{s}-1)N_{r}^{RF}N_{t}^{RF} + N_{r}N_{t} \big)$.
\item Single-timescale DNN: The single timescale DNN applies the long-term DNN to update both analog and digital precoder/combiner in each time slot. Hence, the number of signaling bits in each time slot is the dimension of vector $\mathbf{q}$, i.e., $B(B\ll B_{c}N_{r}N_{t})$, and that over a superframe is given by $Q_{s} = T_{f}T_{s}B$.  
\item Proposed two-timescale DNN: By assuming that $B_{t}$ denotes the dimension of $\mathbf{q}_{eq}$, then we have $B_{t}<B$. Hence, the number of signaling bits of the proposed two-timescale DNN within a superframe is given by $Q_{t} = T_{f}\big((T_{s}-1)B_{t}+B \big)$.
\end{itemize}

Based on the above results, it is readily seen that the proposed two-timescale DNN scheme significantly reduces the signaling overhead compared to other existing schemes.

\subsection{Extension to OFDM Systems}
In this subsection, we introduce how to extend the proposed two-timescale DNN to wideband mmWave OFDM systems. Three key issues need to be considered for the extension \cite{HybridDNN}:
\begin{itemize}
\item  In OFDM systems, the digital precoder and combiner can be designed independently for different subcarriers while the analog precoder and combiner must be shared by all subcarriers.  
	
\item It is important to maintain the architecture of the DNN, i.e., the number of neurons in each layer and the number of layers in the DNN. 
	
\item Since the number of subcarriers are generally large in OFDM systems, the training time of the DNN should not increase with the number of subcarriers. 
\end{itemize}
The signal transmission model is related to the subcarrier and the detected signal of the $k$-th subcarrier is given by
\begin{equation} 
\mathbf{r}[k] = \mathbf{W}^{H}_{BB}[k] \mathbf{W}^{H}_{RF} \mathbf{H}[k] \mathbf{F}_{RF} \mathbf{F}_{BB}[k] \mathbf{s} + \mathbf{W}^{H}_{BB}[k] \mathbf{W}^{H}_{RF} \mathbf{n},
\end{equation}
where $k\in \mathcal{K}\triangleq \{1, 2, \cdots, K\}$ denotes the index of OFDM subcarriers.

\subsubsection{Pilot Training and CSI feedback in the Long-Term DNN}
To estimate the full CSI matrix $\mathbf{H}[k]$, the TX sends the training pilot matrix $\tilde{\mathbf{X}}[k] \in \mathbb{C}^{N_{t}^{RF} \times L}$  modulated by the analog precoder $\tilde{\mathbf{F}}_{RF}\in \mathbb{C}^{N_t \times N^{RF}_{t}}$. Subsequently, the received pilot signal matrix processed by the analog combiner $\tilde{\mathbf{W}}_{RF}\in \mathbb{C}^{N_r \times N^{RF}_{r}}$ is expressed as 
\begin{equation}
\tilde{\mathbf{Y}}[k] = \tilde{\mathbf{W}}_{RF}^{H} \mathbf{H}[k] \tilde{\mathbf{F}}_{RF}\tilde{\mathbf{X}}[k] + \tilde{\mathbf{N}}[k],
\end{equation}
where $\tilde{\mathbf{N}}[k]=\tilde{\mathbf{W}}_{RF}^{H}\mathbf{N}[k]$, and $\mathbf{N}[k]\in \mathbb{C}^{N_{r} \times L}$ denotes an AWGN matrix. 
To model the pilot training process and find the optimal pilots for the estimation of $\mathbf{H}[k]$, the input and output of this DNN are $\mathbf{H}[k]$ and $\tilde{\mathbf{Y}}[k]$, respectively, and the trainable parameters are $\{\tilde{\mathbf{X}}[k], \tilde{\mathbf{F}}_{RF}, \tilde{\mathbf{W}}_{RF}\}$.  

The RX estimates the CSI matrix $\mathbf{H}[k]$ based on the received pilot signal matrix $\tilde{\mathbf{Y}}[k]$. Subsequently, the RX extracts the useful information and feeds back that information as $B[k]$ bits to the TX for hybrid precoding. These two steps can be represented by a $R$-layer DNN, where the feedback bits of the RX are given by
\begin{equation}
\mathbf{q}[k] \!=\! \textrm{sgn} \big( \! \mathbf{W}_{R}\sigma_{R-1}\big( \cdots \sigma_{1} \big( \! \mathbf{W}_{1}\bar{\mathbf{y}}[k]+\mathbf{b}_{1}  \big) \cdots \! \big) + \mathbf{b}_{R} \! \big), 
\end{equation}
where $\mathbf{q}[k]\in \{\pm 1\}^{B[k]}$, $\tilde{\mathbf{y}}[k] \triangleq \textrm{Vec}(\tilde{\mathbf{Y}}[k])$ denotes the vectorization of matrix $\tilde{\mathbf{Y}}[k]$, and the input of DNN is the real representation of $\tilde{\mathbf{y}}[k]$, i.e., $\bar{\mathbf{y}}[k] \triangleq [\Re(\tilde{\mathbf{y}}[k]^{T}), \Im(\tilde{\mathbf{y}}[k]^{T})]^{T}$.  

\subsubsection{Pilot Training and CSI feedback in the Short-Term DNN}
To estimate the low-dimensional equivalent CSI matrix $\mathbf{H}_{eq}[k]$, the TX sends the training pilot matrix $\tilde{\mathbf{X}}_{eq}[k]\in \mathbb{C}^{N_{t}^{RF} \times L}$ and the received pilot signal matrix at the RX is given by 
\begin{equation}
\tilde{\mathbf{Y}}_{eq}[k] = \mathbf{H}_{eq}[k]\tilde{\mathbf{X}}_{eq}[k] + \tilde{\mathbf{N}}_{eq}[k], 
\end{equation}
where $\mathbf{H}_{eq}[k] = \mathbf{W}_{RF}^{H} \mathbf{H}[k] \mathbf{F}_{RF}$, $\tilde{\mathbf{N}}_{eq}[k] = \mathbf{W}_{RF}^{H}\mathbf{N}[k]$, and $\mathbf{N}[k]\in \mathbb{C}^{N_{r} \times L}$ denotes an AWGN matrix. 
To model the pilot training process for the estimation of $\mathbf{H}_{eq}[k]$, the input and output of this DNN are $\mathbf{H}_{eq}[k]$ and $\tilde{\mathbf{Y}}_{eq}[k]$, respectively, and its trainable parameter is $\tilde{\mathbf{X}}_{eq}[k]$. 

The RX estimates $\mathbf{H}_{eq}[k]$ based on the received pilot matrix $\tilde{\mathbf{Y}}_{eq}[k]$ and extracts useful information for feedback with $B_{eq}[k]$ bits. These two steps can be represented by a $R_{eq}$-layer FC DNN and the feedback bits is given by
\begin{equation}
\mathbf{q}_{eq}[k] \!=\! \textrm{sgn} \big( \! \mathbf{W}_{R_{eq}}\sigma_{R_{eq}-1}\big( \! \cdots \sigma_{1} \big( \! \mathbf{W}_{1}\bar{\mathbf{y}}_{eq}[k]+\mathbf{b}_{1}  \big) \cdots \! \big) \!+\! \mathbf{b}_{R_{eq}} \! \big), 
\end{equation}
where $\mathbf{q}_{eq}[k]\in \{\pm 1\}^{B_{eq}[k]}$, $\tilde{\mathbf{y}}_{eq}[k]\triangleq \textrm{Vec} (\tilde{\mathbf{Y}}_{eq}[k])$ denotes the vectorization of matrix $\tilde{\mathbf{Y}}_{eq}[k]$, and the input of DNN is the real representation of $\tilde{\mathbf{y}}_{eq}[k]$, i.e., $\bar{\mathbf{y}}_{eq}[k] \triangleq [\Re(\tilde{\mathbf{y}}^{T}_{eq}[k]), \Im(\tilde{\mathbf{y}}^{T}_{eq}[k])]^{T}$. 

\subsubsection{Hybrid Precoder and Combiner Design in the Long-Term DNN}
The TX collects the feedback bits $\mathbf{q}[k]$ to recover the full CSI matrix $\hat{\mathbf{H}}[k]$. Then, the TX designs the hybrid precoder and combiner based on the recovered full CSI matrix with a DNN. Note that the analog $\mathbf{F}_{RF}$ and $\mathbf{W}_{RF}$ should be shared by all subcarriers, we input $\dot{\mathbf{H}}$ to AP-NN and AC-NN to generate $\bm{\varphi}_{F} \in \mathbb{R}^{N_{t}N^{RF}_{t}\times 1 }$ and $\bm{\varphi}_{W}\in \mathbb{R}^{N_{r}N^{RF}_{r}\times 1 }$, respectively. There are two methods for the choice of $\dot{\mathbf{H}}$: (i) the estimated CSI matrix of a given subcarrier \cite{HybridDNN}, e.g., the $l$-th subcarrier $\hat{\mathbf{H}}[l]$; (ii) the average of the estimated CSI matrix of all the subcarriers, i.e., $\sum_{k=1}^{K} \hat{\mathbf{H}}[k]$. 
With $\bm{\varphi}_{F}$ and $\bm{\varphi}_{W}$, $\mathbf{F}_{RF}$ and $\mathbf{W}_{RF}$ can be generated based on \eqref{PhaseShifter} and \eqref{FRF}. Then, $\mathbf{F}_{RF}$ and $\mathbf{W}_{RF}$ along with $\hat{\mathbf{H}}[k]$ are employed to generate a low-dimensional equivalent CSI as
\begin{equation}
\hat{\mathbf{H}}_{eq}[k]=\mathbf{W}_{RF}^{H}\hat{\mathbf{H}}[k]\mathbf{F}_{RF}. 
\end{equation}
Afterwards, $\hat{\mathbf{H}}_{eq}[k]\in \mathbb{C}^{N^{RF}_{r}\times N^{RF}_{t}}$ is input into the DP-NN and DC-NN, the outputs of which are $\{ \bar{\mathbf{w}}_{BB,re}[k],\bar{\mathbf{w}}_{BB,im}[k] \}$ and $\{ \bar{\mathbf{f}}_{BB,re}[k], \bar{\mathbf{f}}_{BB,im}[k] \}$, respectively. Then, $\mathbf{W}_{BB}[k]$ and $\mathbf{F}_{BB}[k]$ are computed as
\begin{equation} 
\begin{aligned}
&\mathbf{W}_{BB}[k]=\mathcal{J}_{v\rightarrow m}(\bar{\mathbf{w}}_{BB,re}[k] + j\bar{\mathbf{w}}_{BB,im}[k]), \\
&\bar{\mathbf{F}}_{BB}[k]=\mathcal{J}_{v\rightarrow m}(\bar{\mathbf{f}}_{BB,re}[k] + j\bar{\mathbf{f}}_{BB,im}[k]).
\end{aligned}
\end{equation}
Finally, $\bar{\mathbf{F}}_{BB}[k]$ is normalized to satisfy the power constraint. 

\subsubsection{Digital Precoder and Combiner Design in the Short-Term DNN}
The TX collects the feedback bits $\mathbf{q}_{eq}[k]$ to recover the low-dimensional equivalent CSI $\hat{\mathbf{H}}_{eq}[k]$. Then, the TX designs the digital precoder $\mathbf{F}_{BB}[k]$ and combiner $\mathbf{W}_{BB}[k]$ based on the recovered equivalent CSI matrix $\hat{\mathbf{H}}_{eq}[k]$ with a DNN, while the analog precoder $\mathbf{F}_{RF}$ and combiner $\mathbf{W}_{RF}$ are fixed. 

\subsubsection{Training Process}
Compared to the training sample with the input tuple $\{ \mathbf{H}, \mathbf{n}, \mathbf{S}_{b} \}$, we modify the input tuple as $\{ \dot{\mathbf{H}}, \mathbf{H}[k], \mathbf{n}, \mathbf{S}_{b} \}$. Note that the training process of pilot training DNN and CSI feedback DNN is the same as the single-carrier system. In the following, we introduce how to train the long-term hybrid precoding DNN and the short-term digital precoding DNN can be trained similarly.
When inputting each training sample into the hybrid precoding DNN, $\dot{\mathbf{H}}$ will be used to generate $\mathbf{F}_{RF}$ and $\mathbf{W}_{RF}$ via AP-NN and AC-NN. Then, $\mathbf{F}_{RF}$ and $\mathbf{W}_{RF}$ along with $\mathbf{H}[k]$ are used to generate the equivalent CSI of the $k$-th subcarrier $\mathbf{H}_{eq}[k]$, based on which, $\mathbf{F}_{BB}[k]$ and $\mathbf{W}_{BB}[k]$ can be obtained through DP-NN and DC-NN, respectively. On the other hand, $\mathbf{H}[k]$ is also input into the signal flow to act as the fading channel since this training sample is used to simulate the transmission of the $k$-th subcarrier. End-to-end training can then be performed by minimizing the BCE loss between $\mathbf{S}_{b}$ and $\hat{\mathbf{S}}_{b}$. Through training, we can obtain the unified $\mathbf{F}_{RF}$ and $\mathbf{W}_{RF}$ that match the channel of each subcarrier well without complicating the architecture of the proposed two-timescale DNN. 

\section{Simulation Results}  \label{Simulation}
In this section, we verify the effectiveness of the proposed DNN based joint channel acquisition and hybrid precoding algorithm by simulation results. We first present the simulation methodology, followed by the investigation of the convergence in training of the proposed DNN. Then, the proposed algorithm is compared with benchmark approaches.

\subsection{Simulation Setup}
The system configuration is described as follows. We set $N_{t}=64$ and $N_{t}^{RF}=8$ for the TX and $N_{r}=32$ and $N_{r}^{RF}=4$ for the RX. The number of data streams is $N_{s}=4$ and we set $\textrm{SNR} = 10  \textrm{dB}$. The pilot length is set to be $L=28$ and the number of feedback bits is $B=64$. The size of sliding window is set to be $D=3$ and the number of time slots within a frame is $T_{s}=10$. 
We implement the proposed DNN by using the deep learning library ``Pytorch". The ``Adam" optimizer is employed as the training method, with the batch size of $128$ and a learning rate $\eta$ gradually decreasing from $10^{-2}$ to $10^{-5}$. 
To accelerate the convergence speed, each layer is processed by the batch normalization layer and drop-out technique. In the training stage, we slowly increase the annealing parameter of the sigmoid-adjusted ST in \eqref{SigmoidST} as $\alpha^{i}=2+0.2i$, where $i$ denotes the index of epoch and each epoch consists of $200$ mini-batches. 

We employ the widely used narrowband mmWave clustered channel \cite{MaGiQ}, which consists of $N_{cl}$ clusters with $N_{ray}$ propagating rays. The CSI matrix can be expressed as
\begin{equation} \label{channel}
\mathbf{H} = \sqrt{\dfrac{N_{t}N_{r}}{N_{cl}N_{ray}}}\sum\limits_{i=1}^{N_{cl}}\sum\limits_{l=1}^{N_{ray}} \alpha_{il} \mathbf{a}_{r}(\phi_{il}^{r}) \mathbf{a}_{t}^{H}(\phi_{il}^{t}),
\end{equation}
where $\alpha_{il}\sim \mathcal{CN}(0,\sigma_{\alpha}^{2})$ is the complex gain of the $l$-th ray in the $i$-th cluster, $\phi_{il}^{r}$ and $\phi_{il}^{t}$ denote the azimuth AoA and AoD at the RX and TX for the $l$-th ray in the $i$-th cluster, respectively. The $\mathbf{a}_{r}(\phi_{il}^{r})$ and $\mathbf{a}_{t}(\phi_{il}^{t})$ represent the receive and transmit array response vectors, respectively. For a uniform linear array with $N$ antenna elements and an azimuth angle of $\phi$, the response vector can be written as
\begin{equation}
\mathbf{a}(\phi) \!=\! \frac{1}{\sqrt{N}}\big[1, e^{-j 2\pi \frac{d}{\lambda} \sin(\phi)}, \cdots , e^{-j 2\pi \frac{d}{\lambda}(N-1) \sin(\phi)}  \big]^{T},
\end{equation}
where $d$ and $\lambda$ denote the distance between the adjacent antennas and carrier wavelength, respectively. 
We select $N_{cl}=3$ clusters and $N_{ray}=4$ rays in each cluster, where $\alpha_{il}\sim \mathcal{CN}(0,1)$, $\phi_{il}^{r}\sim \mathcal{U}(-\frac{\pi}{2}, \frac{\pi}{2})$, and $\phi_{il}^{t}\sim \mathcal{U}(-\frac{\pi}{2}, \frac{\pi}{2})$. 

Based on \eqref{channel}, we introduce the CSI mismatch $\textrm{exp}(j2\pi f_{d} \tau \textrm{cos}(\phi_{il}^{r}))$ with the CSI delay $\tau$ and the maximum Doppler shift $f_{d}$ for the channel model. Thus, the actual channel matrix is modeled as \cite{TwotimeChannel}
\begin{equation} \label{ChannelDelay}
\mathbf{H} \!=\! \sqrt{\dfrac{N_{t}N_{r}}{N_{cl}N_{ray}}} \! \sum\limits_{i=1}^{N_{cl}} \! \sum\limits_{l=1}^{N_{ray}} \! \alpha_{il} \mathbf{a}_{r}(\phi_{il}^{r}) \mathbf{a}_{t}^{H}(\phi_{il}^{t}) \! \times \! \textrm{exp}(j2\pi f_{d} \tau \textrm{cos}( \! \phi_{il}^{r} \! ) \! ).
\end{equation}
Note that the CSI delay is proportional to the number of CSI feedback bits \cite{TwotimeChannel} as $\dfrac{\tau_{t}}{\tau_{s}}=\dfrac{Q_{t}}{Q_{s}}$, where $\tau_{t}$ and $\tau_{s}$ denote the CSI delay of the two-timescale algorithm and the single-timescale algorithm, respectively. In the simulation, we set $\tau_{s}=1$ ms.

As for benchmarks, we adopt two algorithms for hybrid precoding: (i) the iterative optimization algorithm (OPT) proposed in \cite{MaGiQ} and (ii) the heuristic channel matching algorithm (CMA) developed in \cite{ChannelMatch}. The optimal Lloyd-Max algorithm is employed to quantize the channel parameters and the orthogonal matching pursuit (OMP) \cite{OMP} is applied to estimate the CSI matrix.
In particular, we compare the performance of the following methods:
\begin{itemize}
\item Proposed two-timescale DNN: The proposed DNN in the two-timescale fashion with the long-term and short-term DNNs, where the CSI mismatch caused by the delay is considered.
\item Single-timescale DNN: The proposed DNN in the single-timescale fashion that only employs the long-term DNN and updates both the analog and digital precoder/combiner in each time slot.
\item  OPT (CMA): The full CSI matrix $\mathbf{H}$ is perfectly known at the TX and the OPT (CMA) is employed to design the hybrid precoding matrices. 
\item OPT (CMA)/Lloyd: The RX has perfect knowledge about its channel parameters, i.e., $\{ \Re(\alpha_{il}), \Im(\alpha_{il}), \phi_{il}^{r}, \phi_{il}^{t}, \forall i, l \}$. It aims to transmit these parameters to the TX by sending the quantized version of the channel parameters based on the Lloyd-Max algorithm, over an error-free $B$-bits finite-capacity feedback link \cite{FeedbackFDD}. Note that each channel parameter is allocated to $\frac{B}{4N_{cl}N_{ray}}$ quantization bits. By employing the channel model in \eqref{channel}, the TX can reconstruct the estimated CSI matrix $\hat{\mathbf{H}}$, and then the OPT (CMA) is employed to perform hybrid precoding. 
\item OPT (CMA)/OMP: The RX estimates the CSI matrix in the pilot training stage based on the widely-used OMP algorithm and subsequently feeds back the estimated CSI matrix to the TX over an infinite-capacity link. Then, the TX applies the OPT (CMA) to design the hybrid precoding matrices based on the estimated CSI matrix at the RX.
\item  OPT (CMA)/Delay: The OPT (CMA) scheme that takes into account the CSI mismatch caused by the delay.
\item  OPT (CMA)/Lloyd/Delay: The OPT (CMA)/Lloyd scheme that takes into consideration the CSI mismatch caused by the delay.
\item OPT (CMA)/OMP/Delay: The OPT (CMA)/OMP scheme that takes into account the CSI mismatch caused by the delay.
\end{itemize}

\subsection{Convergence and BER Performance}

\begin{figure*}[!t]
	\centering
	\subfloat[]{\centering \scalebox{0.49}{\includegraphics{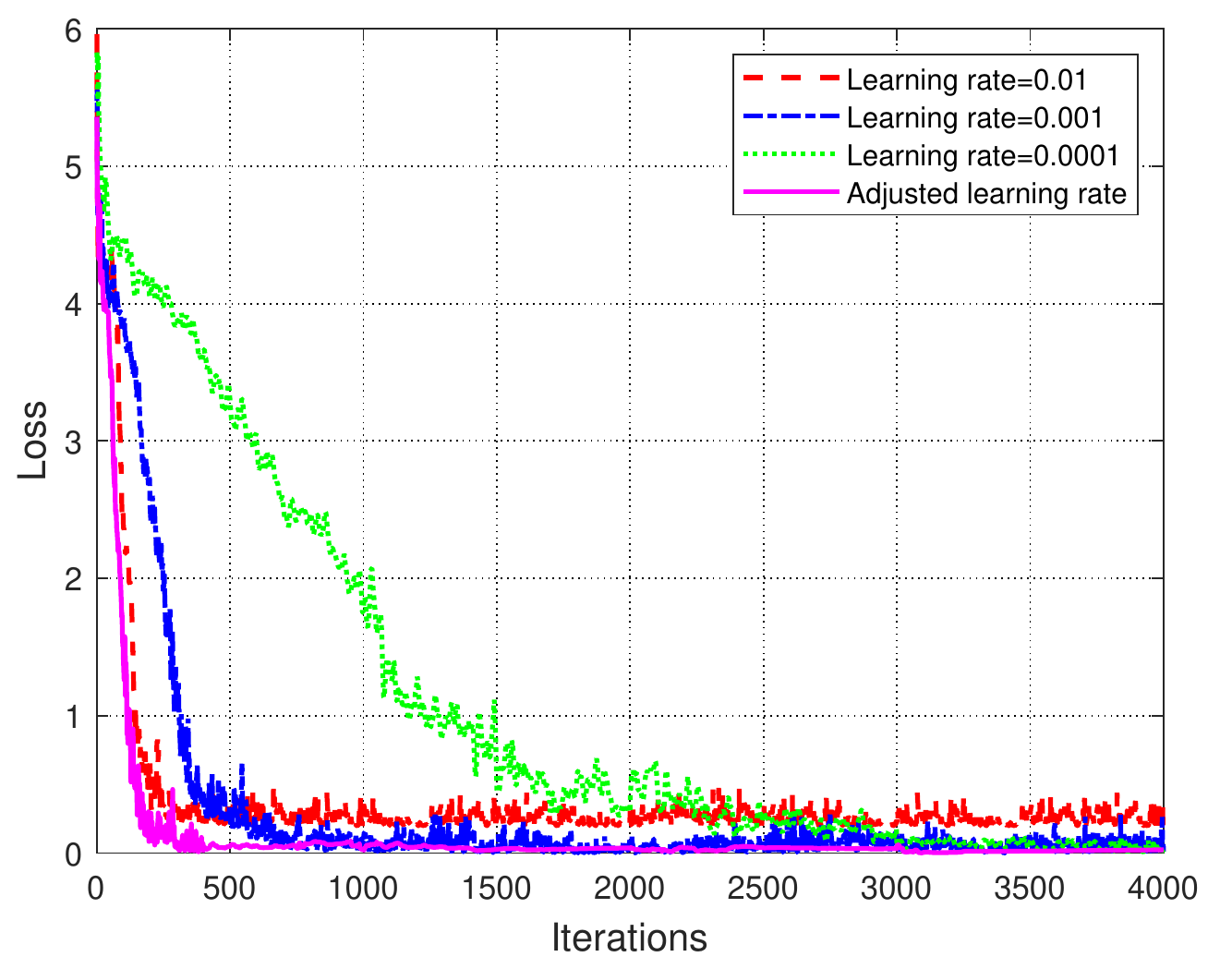}} }
	\subfloat[]{\centering \scalebox{0.49}{\includegraphics{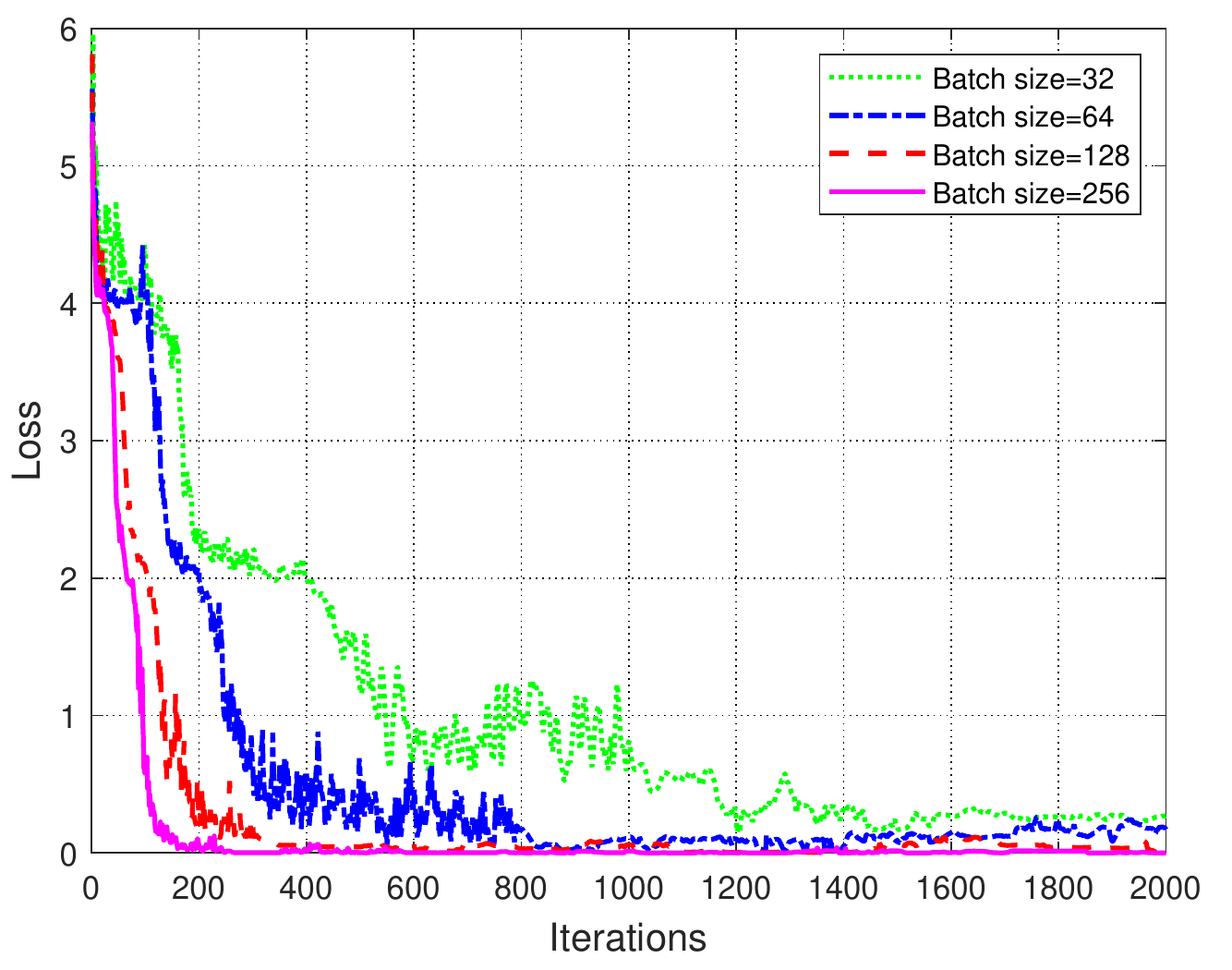}}}	
	\caption{Convergence performance: (a) Learning rate; (b) Batch size.}
	\label{LearningBatch}
\end{figure*}

Fig. \ref{LearningBatch}(a) presents the convergence performance of the loss function, i.e., BCE, with different learning rates. We can see that a smaller learning rate achieves better performance, while a larger learning rate results in faster convergence speed. Note that the adjusted learning rate that progressively decreasing from $10^{-2}$ to $10^{-5}$ achieves satisfactory performance with fast convergence speed.
Fig. \ref{LearningBatch}(b) shows the convergence performance of BCE with different batch sizes. It is observed that a larger batch size leads to more stable convergence.

\begin{figure*}[!t]
	\centering
	\subfloat[]{\centering \scalebox{0.5}{\includegraphics{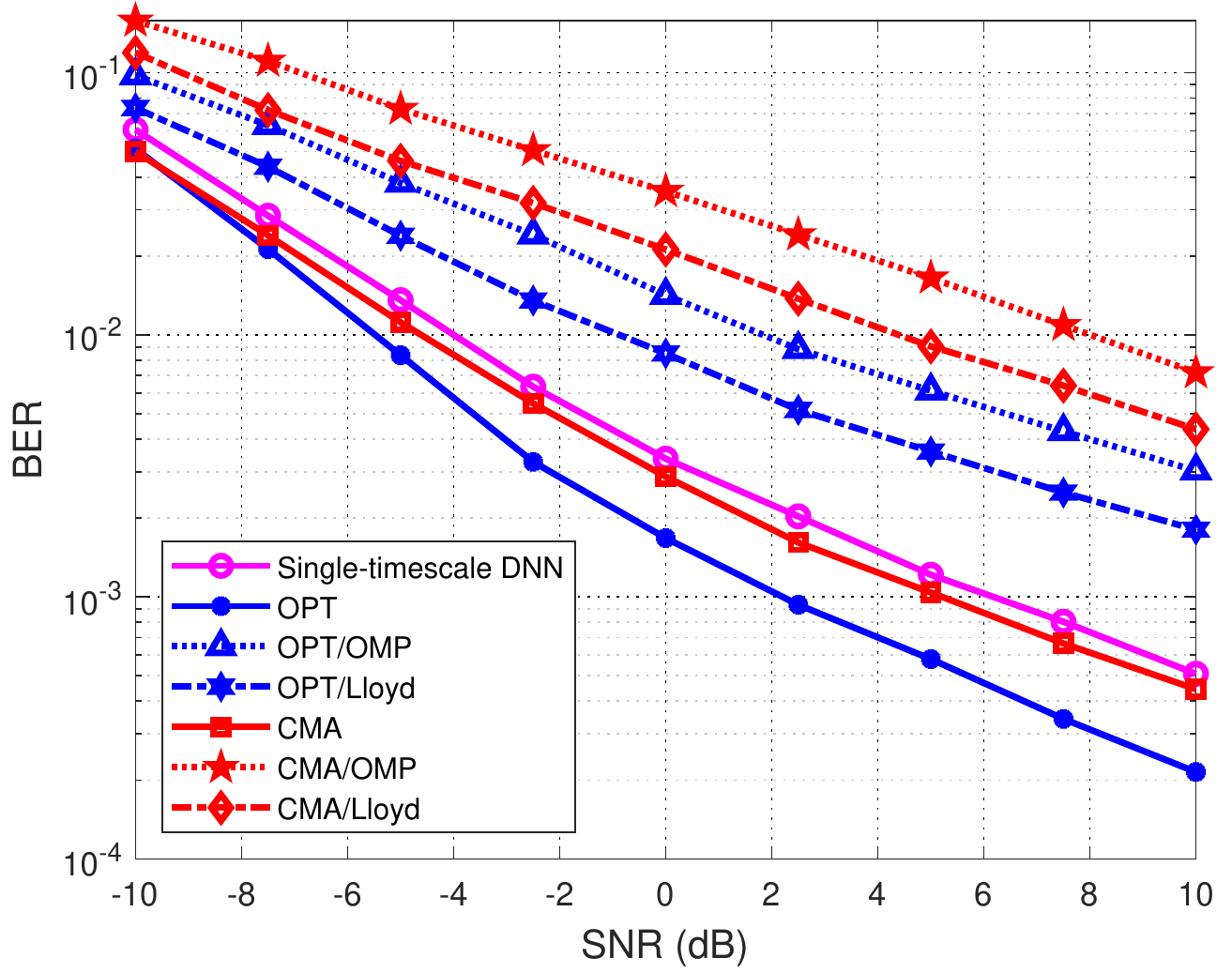}} }
	\subfloat[]{\centering \scalebox{0.5}{\includegraphics{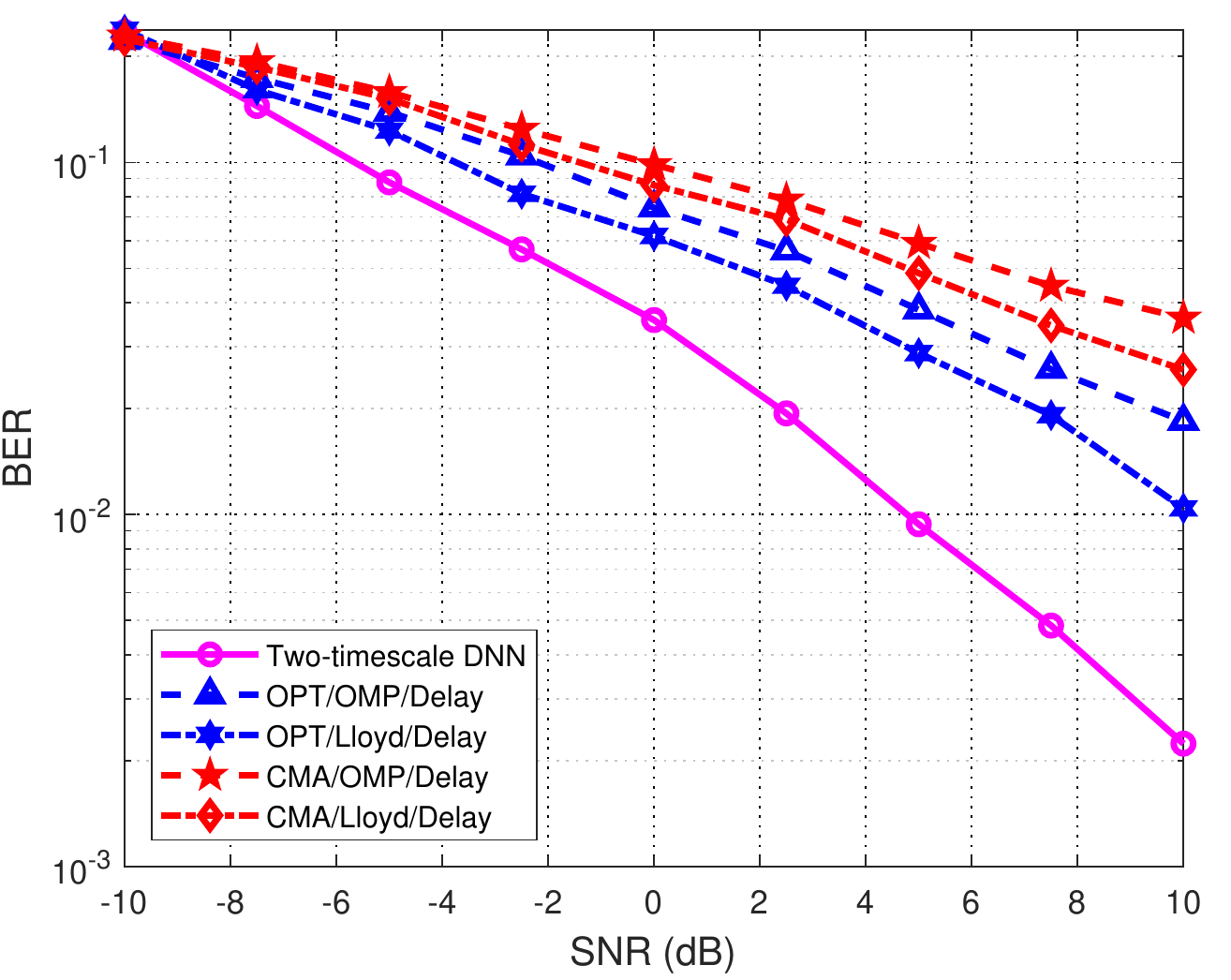}}}	
	\caption{BER performance versus SNR: (a) Proposed DNN in the single-timescale fashion; (b) Proposed two-timescale DNN in the presence of the CSI delay.}
	\label{SNR}
\end{figure*}

\begin{figure*}[!t]
	\centering
	\subfloat[]{\centering \scalebox{0.5}{\includegraphics{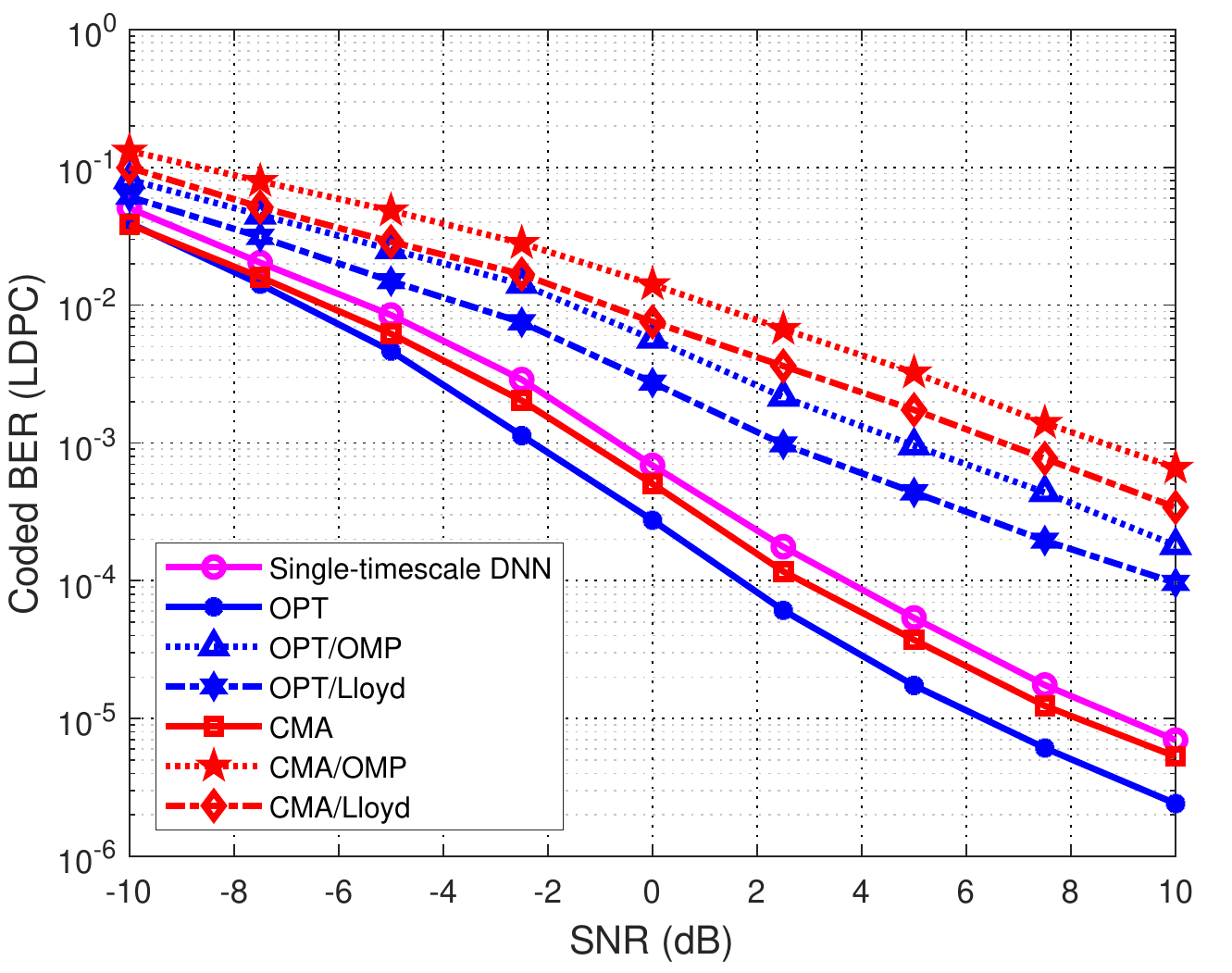}} }
	\subfloat[]{\centering \scalebox{0.5}{\includegraphics{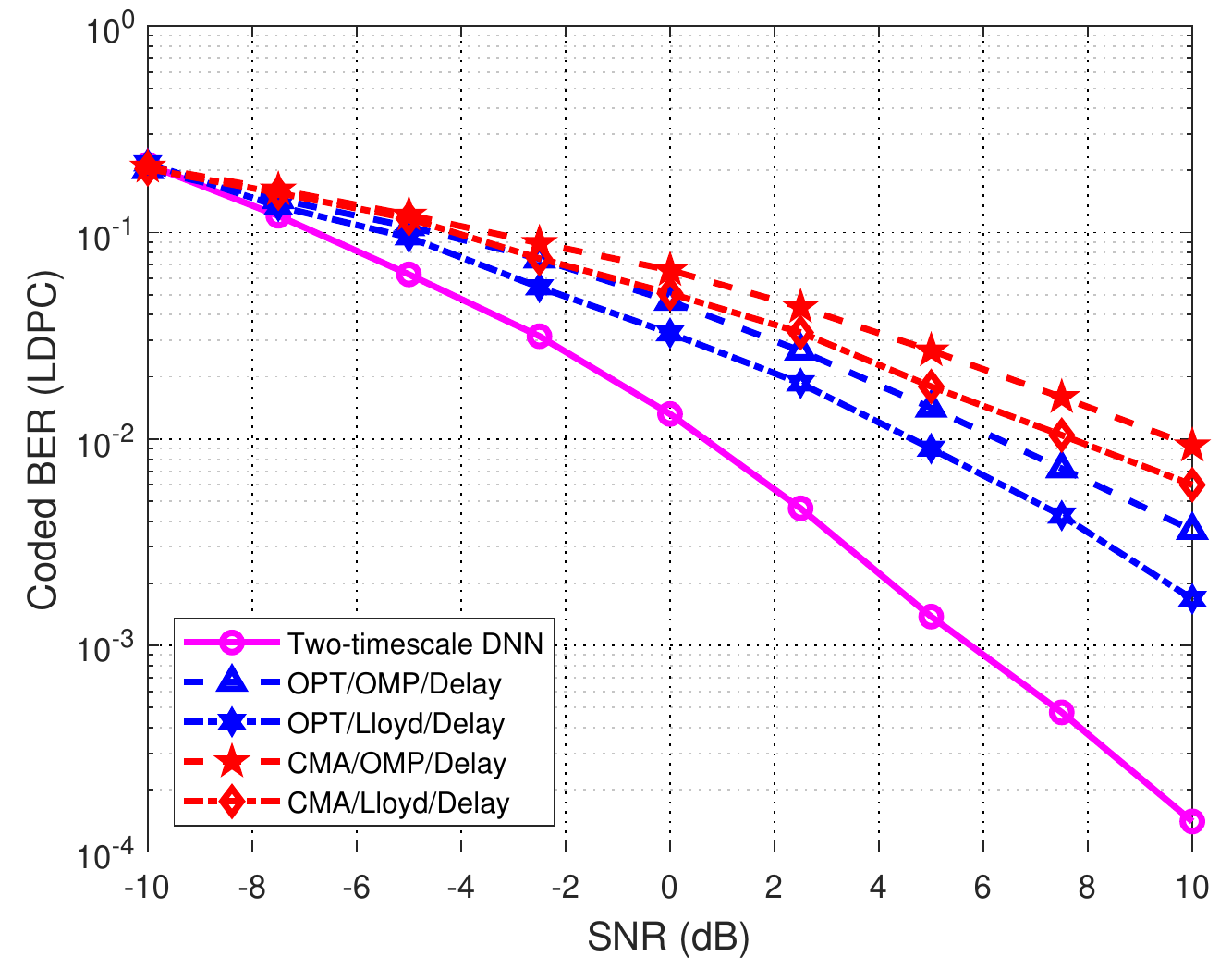}}}	
	\caption{Coded (LDPC) BER performance versus SNR: (a) Proposed DNN in the single-timescale fashion; (b) Proposed two-timescale DNN in the presence of the CSI delay.}
	\label{SNRLDPC}
\end{figure*}

\begin{figure*}[!t]
	\centering
	\subfloat[]{\centering \scalebox{0.5}{\includegraphics{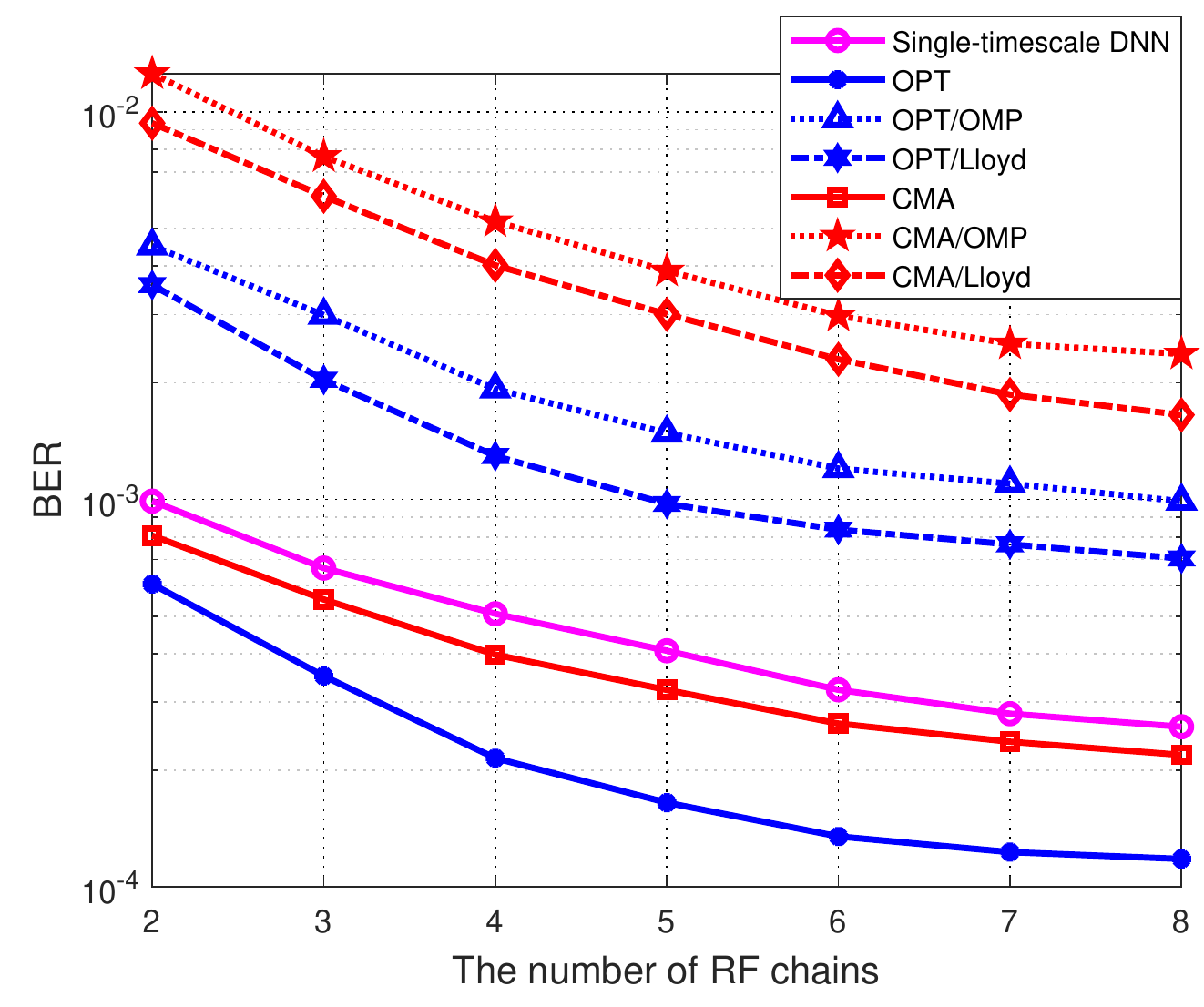}} }
	\subfloat[]{\centering \scalebox{0.5}{\includegraphics{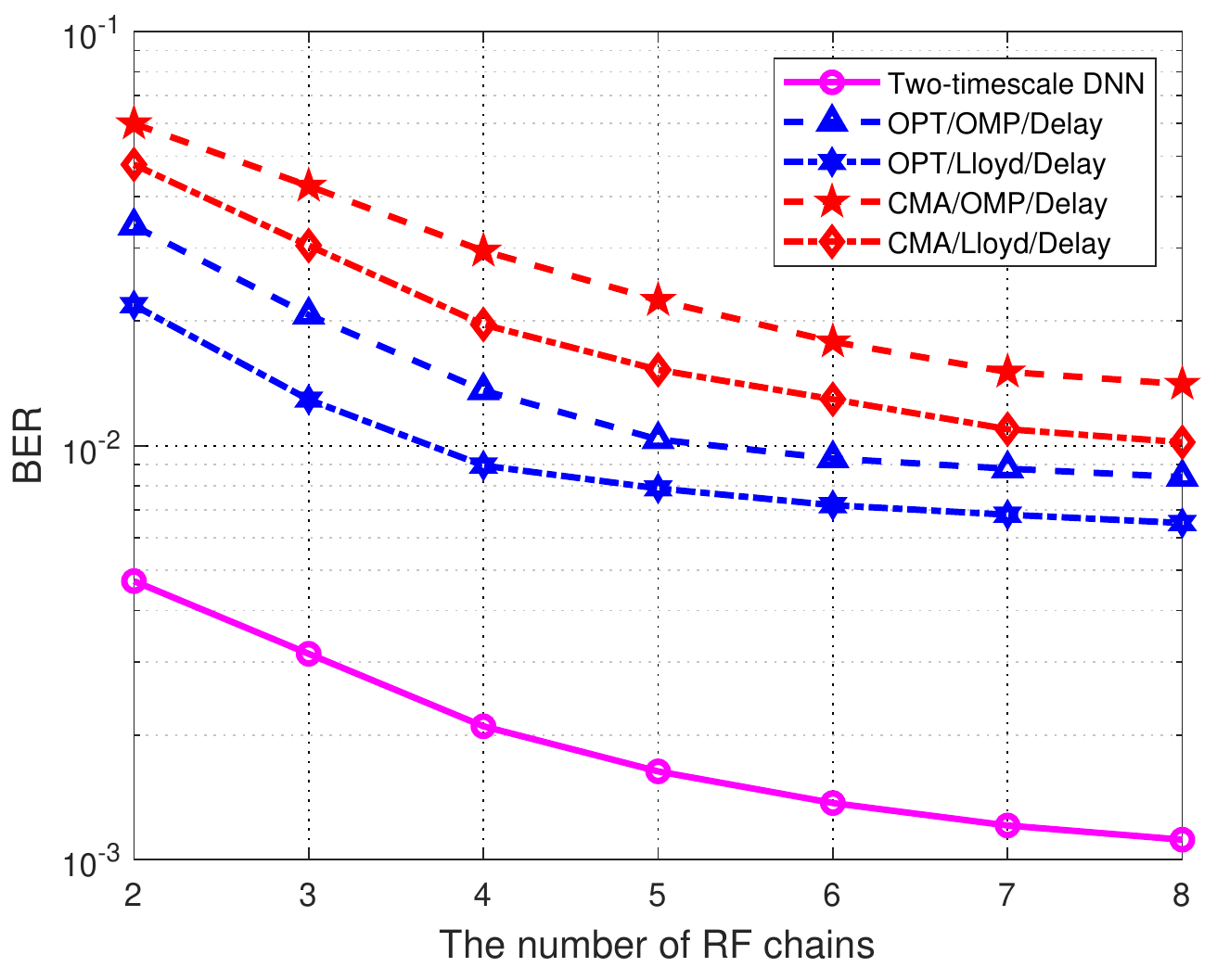}}}	
	\caption{BER performance versus the number of RF chains $N_{RF}$: (a) Proposed DNN in the single-timescale fashion; (b) Proposed two-timescale DNN  in the presence of the CSI delay.}
	\label{RF}
\end{figure*}

\begin{figure*}[!t]
	\centering
	\subfloat[]{\centering \scalebox{0.5}{\includegraphics{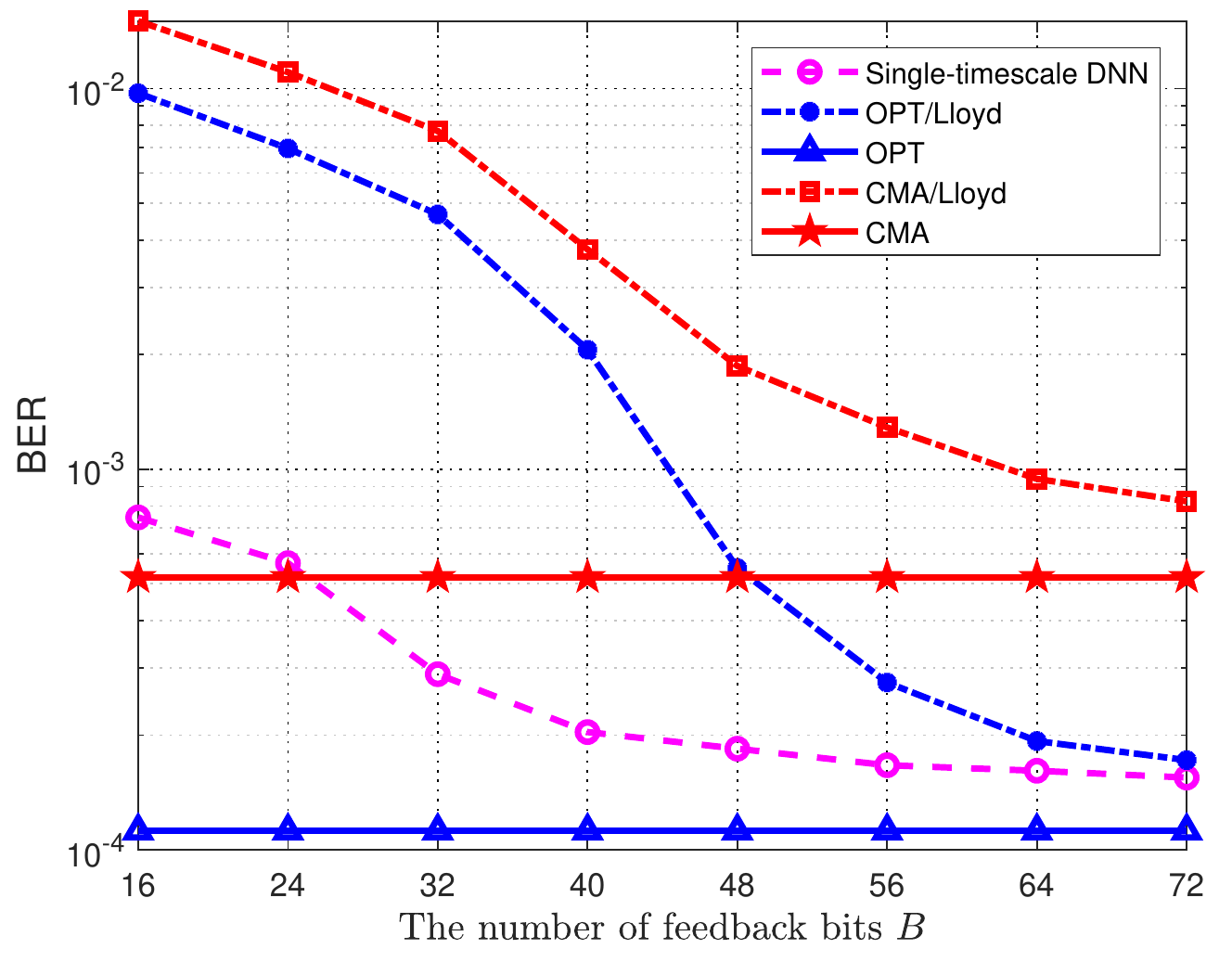}} }
	\subfloat[]{\centering \scalebox{0.5}{\includegraphics{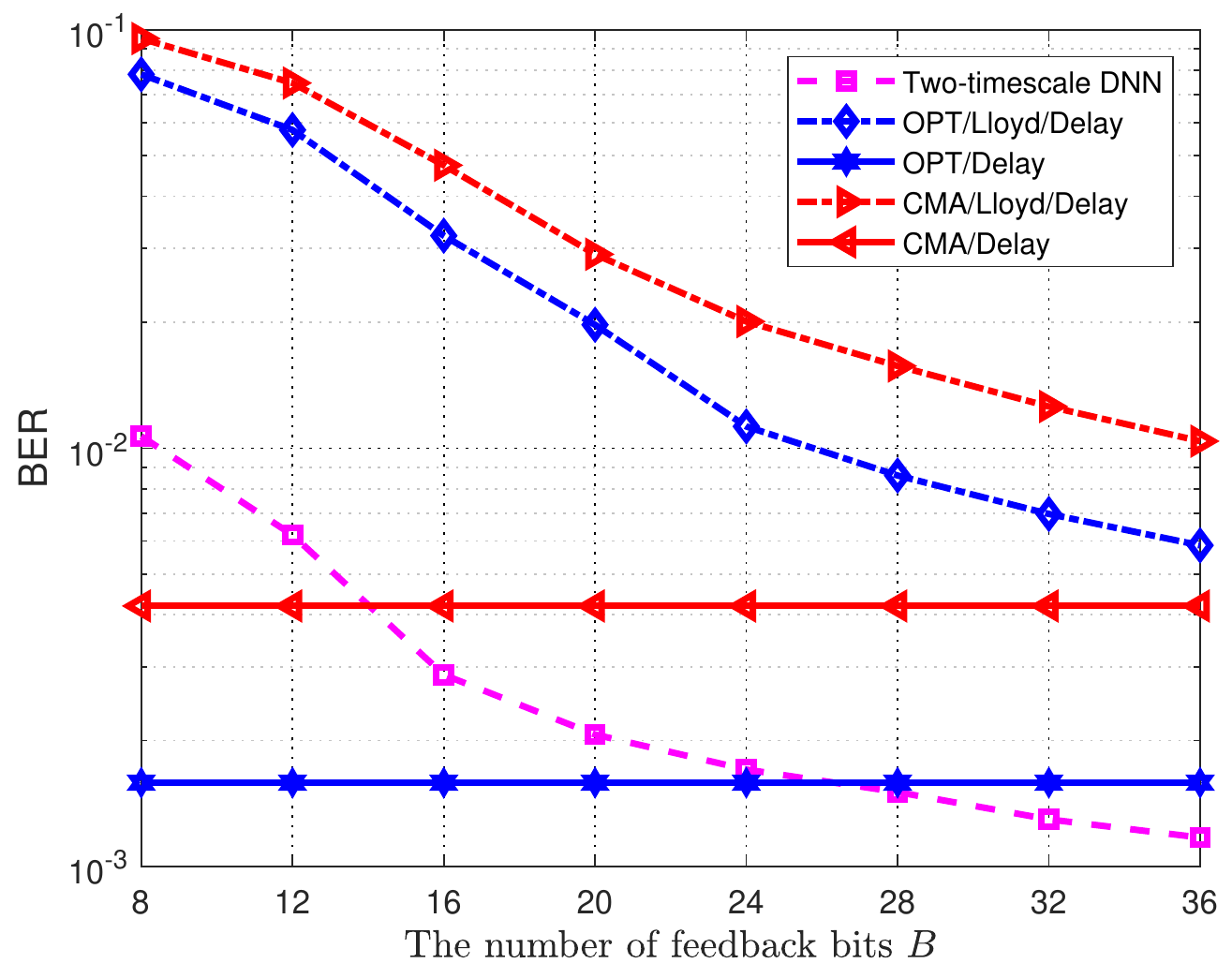}}}	
	\caption{BER performance versus the number of feedback bits $B$: (a) Proposed DNN in the single-timescale fashion; (b) Proposed two-timescale DNN in the presence of the CSI delay.}
	\label{Bits}
\end{figure*}

Fig. \ref{SNR}(a) illustrates the BER performance of the proposed DNN in the single-timescale fashion and the benchmark algorithms for different  values of SNR. We can see that the BER achieved by all the analyzed algorithms decreases monotonically with SNR. The proposed single-timescale DNN outperforms the OPT/Lloyd, OPT/OMP, CMA/Lloyd, and CMA/OMP, where the gap increases with SNR. Thus, the proposed jointly trained single-timescale DNN significantly outperforms the schemes with seperate design of channel estimation, feedback, and hybrid precoding. Moreover, OPT significantly achieves better performance than CMA since it is an iterative optimization algorithm that is guaranteed to find a local optimum, while CMA is a heuristic algorithm. In addition, the BER performance of the single-timescale DNN approaches the lower bound achieved by OPT with perfect CSI and infinite feedback bits. Thus, the proposed single-timescale DNN is indeed an efficient framework for the joint design of pilot training, channel feedback, and hybrid precoding.
 
Fig. \ref{SNR}(b) presents the BER performance of the proposed two-timescale DNN and benchmark algorithms in the presence of CSI delay. We see that the two-timescale DNN significantly outperforms the other benchmarks in terms of BER performance with CSI delay, which verifies the effectiveness of the two-timescale DNN to reduce signaling overhead and CSI mismatch caused by delay.
Fig. \ref{SNRLDPC} presents the coded BER performance of the proposed DNN and the benchmark algorithms, where low-density parity-check (LDPC) codes are employed. Compared with the uncoded BER performance presented in Fig. \ref{SNR}, we see that using an LDPC code significantly improves the BER in high SNR scenarios.

\begin{figure*}[!t]
	\centering
	\subfloat[]{\centering \scalebox{0.5}{\includegraphics{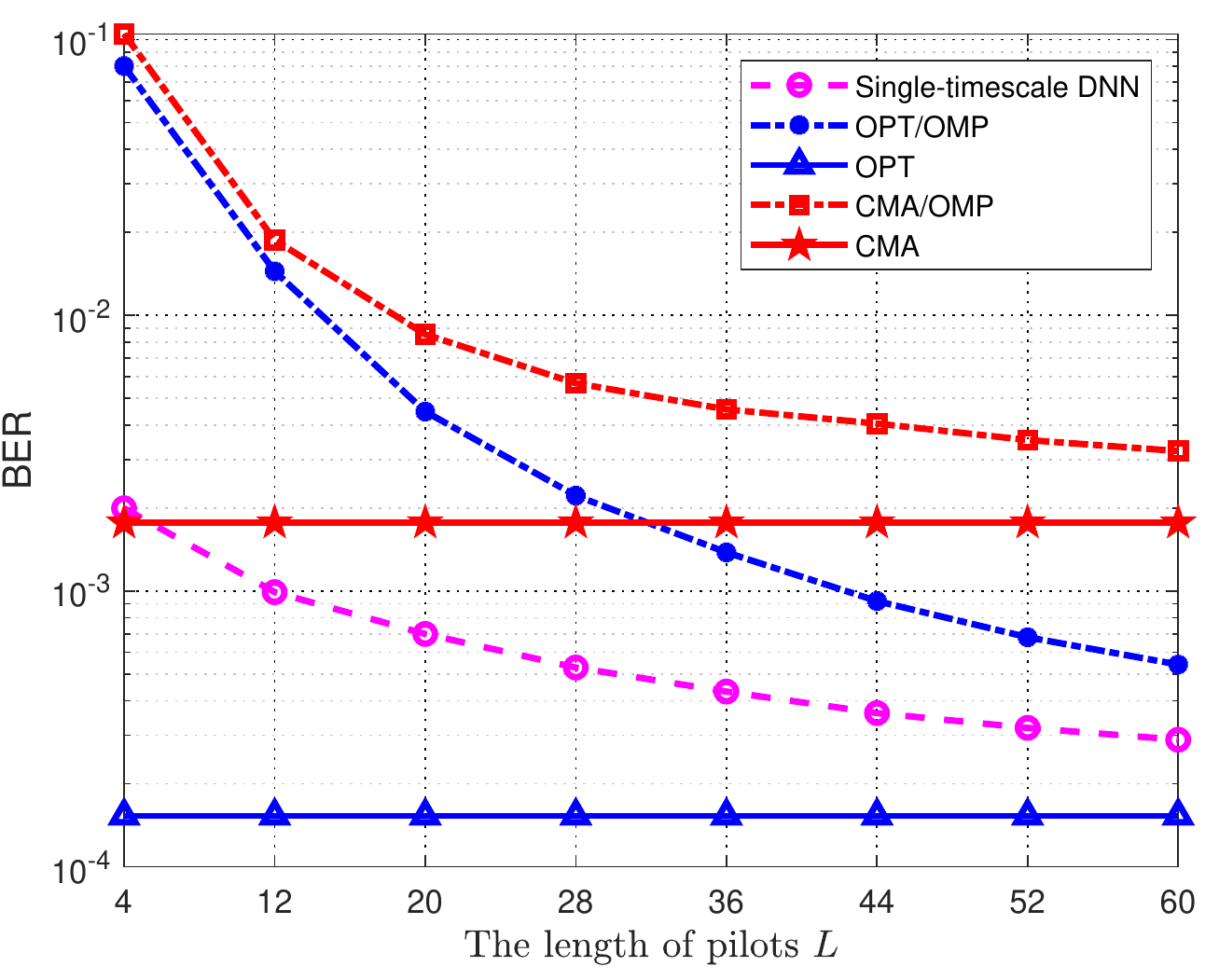}} }
	\subfloat[]{\centering \scalebox{0.5}{\includegraphics{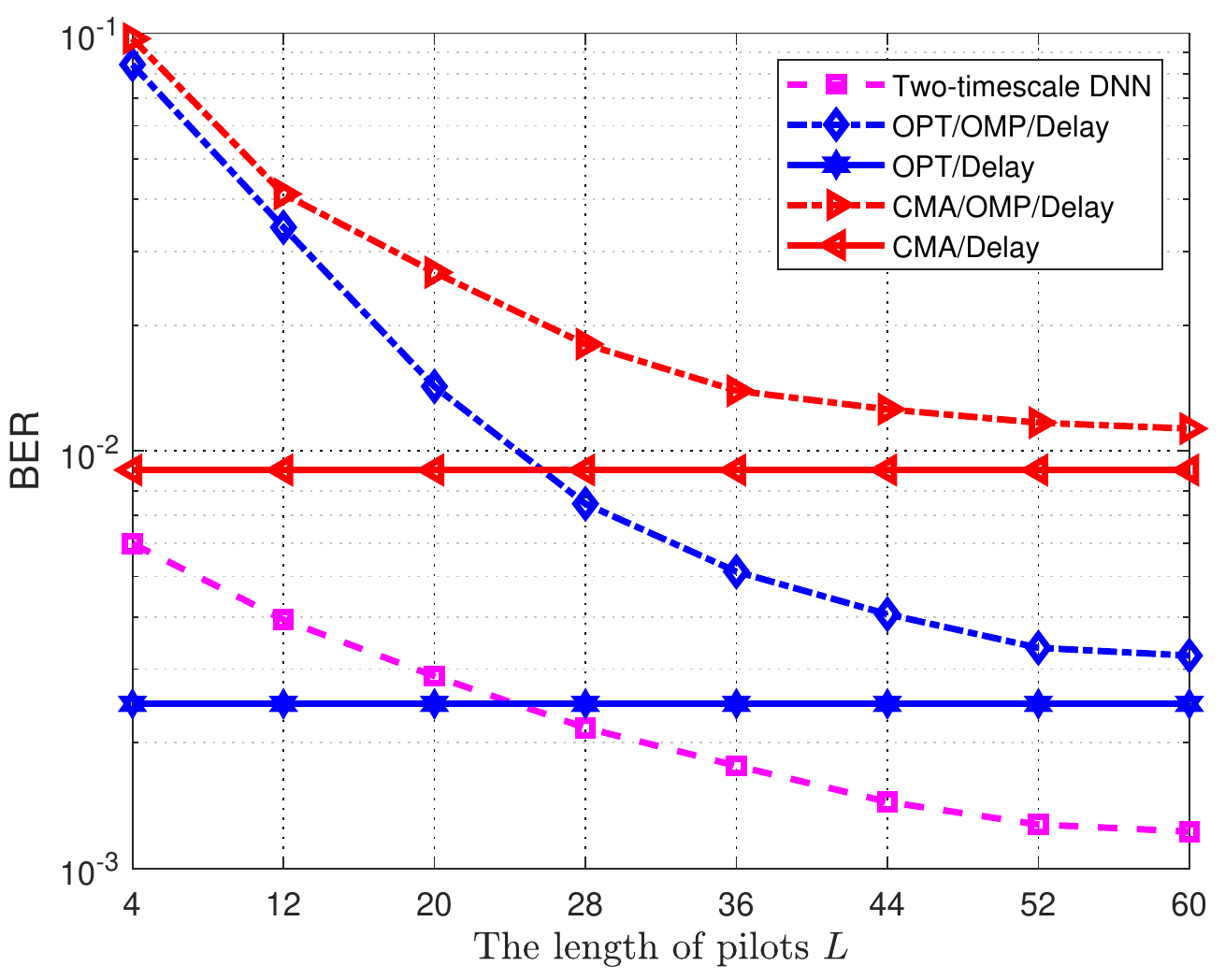}}}	
	\caption{BER performance versus the length of pilots $L$: (a) Proposed DNN in the single-timescale fashion; (b) Proposed two-timescale DNN in the presence of the CSI delay.}
	\label{Pilot}
\end{figure*}

Fig. \ref{RF}(a) shows the BER performance of the proposed DNN in the single-timescale fashion and the benchmark algorithms versus $N_{RF}$. We see that the BER achieved by all the analyzed algorithms decreases monotonically with $N_{RF}$. The proposed single-timescale DNN significantly outperforms OPT/Lloyd, OPT/OMP, CMA/Lloyd, and CMA/OMP, where the gap increases with $N_{RF}$. Moreover, the BER performance of the single-timescale DNN approaches the lower bound achieved by OPT with perfect CSI and infinite feedback bits.
Fig. \ref{RF}(b) presents the BER performance of the proposed two-timescale DNN and the benchmarks versus $N_{RF}$ in the presence of CSI delay. We see that the two-timescale DNN achieves the best BER performance, which further demonstrates the superiority of the proposed DNN-based algorithm.

Fig. \ref{Bits}(a) depicts the BER performance versus $B$ for the single-timescale scenario. We see that the proposed DNN in the single-timescale case outperforms OPT and CMA with the same number of feedback bits $B$ and the gain is significantly large when $B$ is small. This implicitly means that when the number of feedback bits is limited, e.g., $B = 16$, the joint design scheme dramatically outperforms seperate design of the CSI estimation, quantization, and hybrid precoding. Note that the proposed DNN with $B=40$ achieves nearly the same performance as that of OPT with the existing Lloyd-Max CSI quantization method for $B=64$, which shows that our proposed DNN can significantly reduce the number of feedback bits. Furthermore, the proposed DNN provides better performance than CMA with infinite feedback bits. Moreover, it approaches the BER performance achieved by OPT with infinite feedback, which can be regarded as a lower bound. 
Fig. \ref{Bits}(b) shows the BER performance versus $B$ for the two-timescale scenario in the presence of CSI delay. We see that the proposed two-timescale DNN significantly outperforms OPT and CMA with the same number of feedback bits $B$. Furthermore, the proposed DNN with $B=12$ achieves nearly the same performance as that of OPT with $B=36$, which verifies that the proposed two-timescale DNN can reduce the number of feedback bits remarkably. Note that the two-timescale DNN even outperforms OPT with infinite feedback bits when $B>28$ since the two-timescale DNN has stronger robustness against CSI delay. Compared with the single-timescale DNN, the two-timescale DNN can significantly reduce the number of feedback bits since the RX only needs to feed the low-dimensional equivalent CSI $\mathbf{H}_{eq}$ back to the TX.

Fig. \ref{Pilot}(a) shows the BER performance versus the length of pilots $L$ for the single-timescale scenario. It is readily seen that the proposed DNN in the single-timescale fashion outperforms OPT and CMA with the same length of pilots and the gain is obvious when $L$ is small. This demonstrates that when the number of CSI observations is limited, e.g., $L = 12$, joint design is better than seperate design. Note that the proposed DNN with $L=28$ achieves nearly the same performance as that of OPT with conventional OMP channel recovery for $L=60$. This shows that our proposed DNN achieves better BER performance with a reduced number of $L$. Furthermore, the proposed DNN shows better performance than CMA with perfect CSI and approaches the lower bound achieved by OPT with perfect CSI. 
Fig. \ref{Pilot}(b) illustrates the BER performance versus the length of pilots $L$ for two-timescale scenario with CSI delay. We can see that the proposed two-timescale DNN significantly outperforms OPT and CMA with the same pilot length. Furthermore, the proposed DNN with $L=20$ achieves nearly the same performance as that of OPT with $L=60$, which shows that the proposed two-timescale DNN can dramatically reduce the number of $L$. Note that the two-timescale DNN can even outperform OPT with perfect CSI when $L>24$, which verifies the effectiveness of the proposed two-timescale DNN against CSI mismatch caused by delay.

\begin{figure}[t]
\begin{centering}
\includegraphics[width=0.4\textwidth]{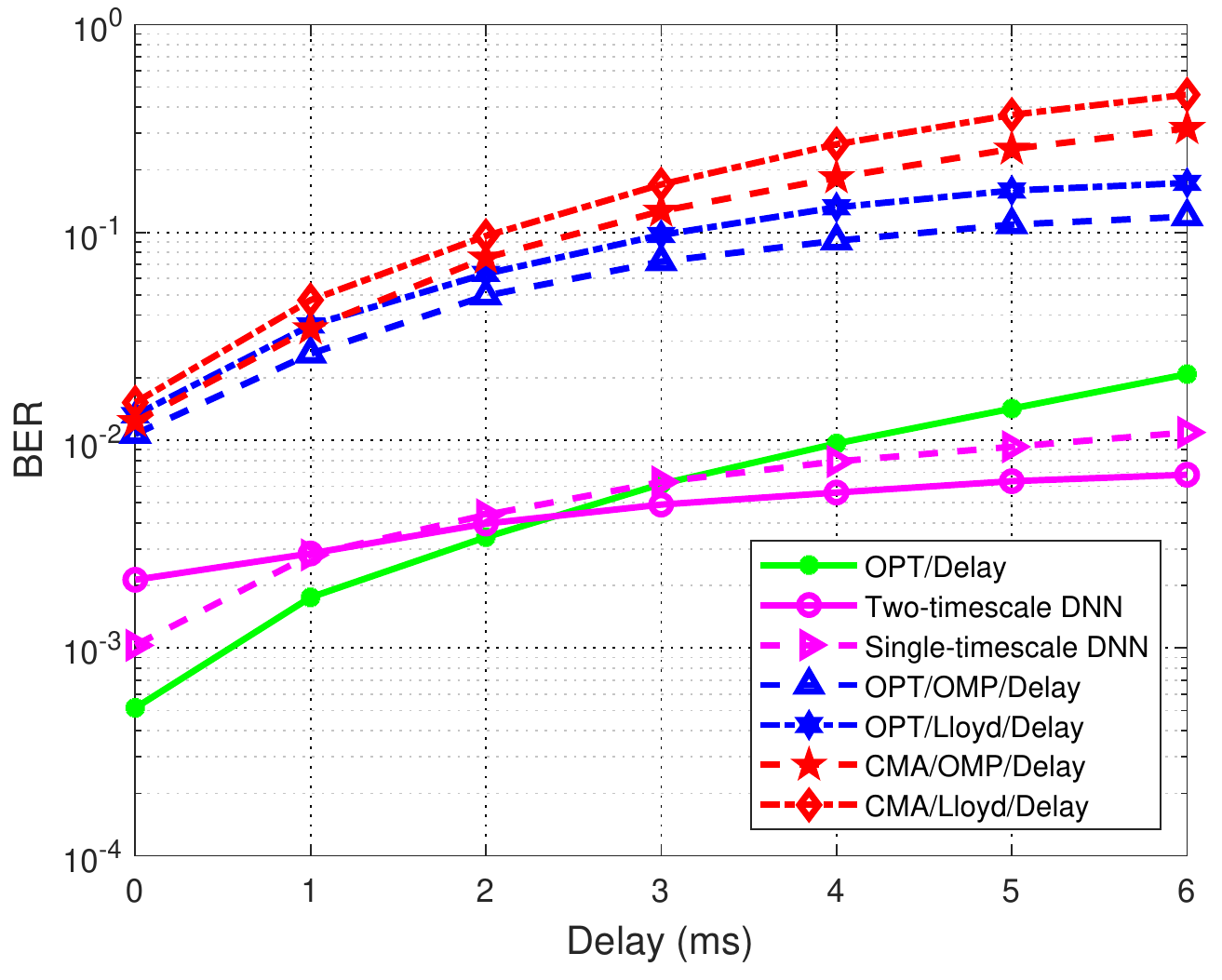}
\par\end{centering}
\caption{BER performance versus the delay $\tau$.}
\label{Delay}
\end{figure}

In Fig. \ref{Delay}, we see that as the delay $\tau$ increases, the performance of the conventional single-timescale precoding algorithms degrades dramatically, while that of the proposed two-timescale DNN changes only slightly due to the large savings of signalling bits. In particular, the two-timescale DNN starts to outperform ``OPT/Delay" when the delay exceeds $3$ ms. These results verify the effectiveness of the proposed algorithm against the CSI errors caused by the delay.

\begin{figure}[t]
\begin{centering}
\includegraphics[width=0.4\textwidth]{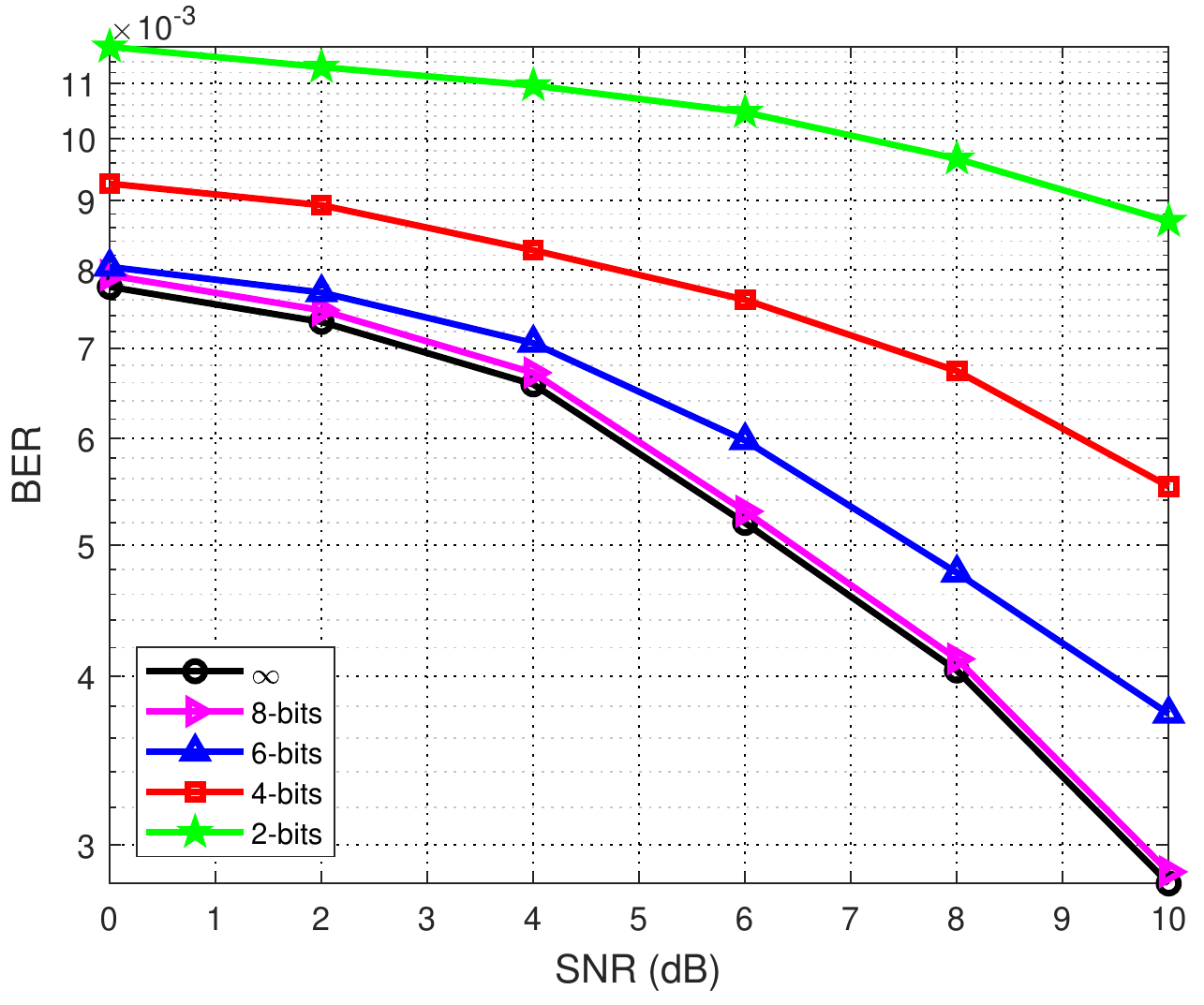}
\par\end{centering}
\caption{BER performance versus the SNR for different numbers of phase shifter quantization bits $Q_{RF}$.}
\label{AnalogBits}
\end{figure}	

\begin{figure}[t]
	\begin{centering}
		\includegraphics[width=0.4\textwidth]{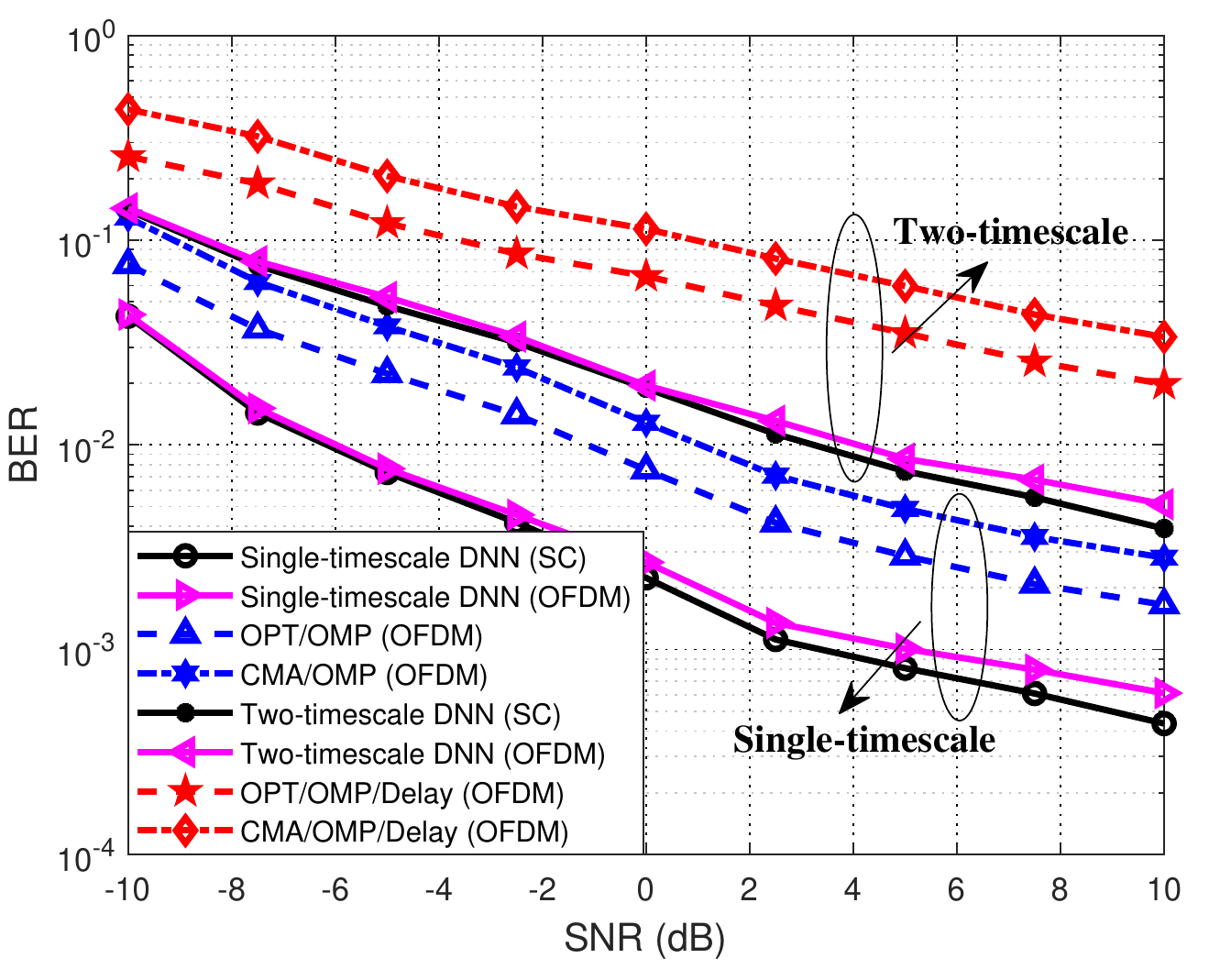}
		\par\end{centering}
	\caption{BER performance in OFDM systems.}
	\label{OFDMSNR}
\end{figure}

Fig. \ref{AnalogBits} presents the BER performance versus SNR for different numbers of phase shifter quantization bits $Q_{RF}$. It can be seen that the performance of the proposed algorithm improves with $Q_{RF}$ as expected. In particular, the performance with $Q_{RF} = 8$ bits can approach the performance with infinite resolution phase shifters.

Fig. \ref{OFDMSNR} verifies that our proposed two-timescale DNN still achieves satisfactory BER performance in OFDM systems. 
To simulate the mmWave wideband and frequency selective MIMO channel in OFDM systems, we employ the clustered delay line (CDL)-B channel model specified in 3GPP R16 \cite{3GPP}. The delay spread and user speed are set as $1$ us and $3$ km/h, respectively. The number of physical resource block (PRB) is $24$ and each PRB contains $12$ subcarriers. Thus, the number of subcarriers is $288$ and the subcarrier space is set as $30$ kHz. 
We can see that the performance achieved by the proposed single-timescale and two-timescale DNNs in OFDM systems approaches that of the single-subcarrier (SC) systems with narrowband mmWave channel, and significantly outperforms the benchmarks.

\begin{figure*}[!t]
	\centering
	\subfloat[]{\centering \scalebox{0.5}{\includegraphics{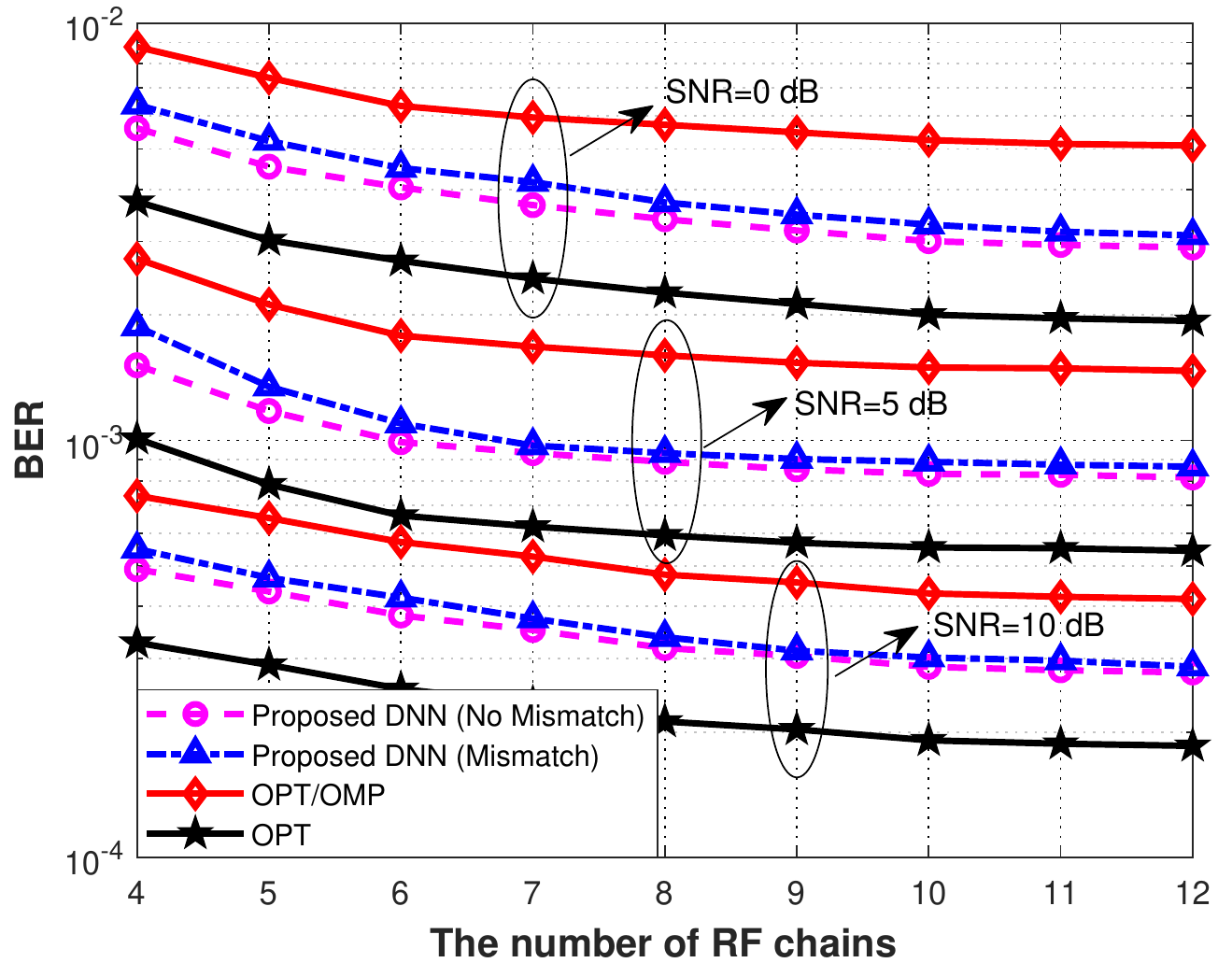}} }
	\subfloat[]{\centering \scalebox{0.5}{\includegraphics{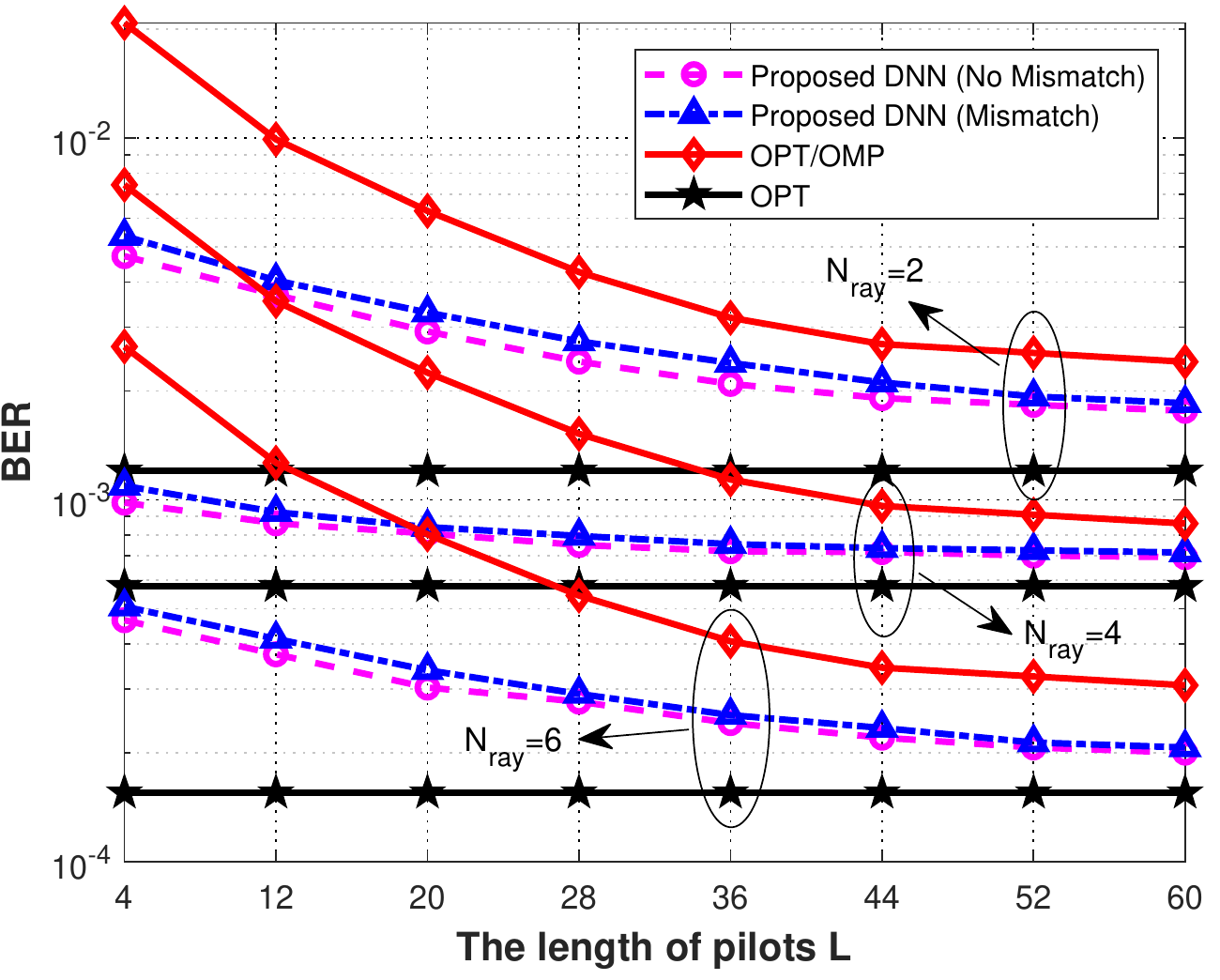}}}	
	\caption{Generalization ability: (a) The number of RF chains $N_{RF}$ and SNR; (b) The length of training pilot $L$ and  the number of propagating rays $N_{ray}$.}
	\label{GeneralRFL}
\end{figure*}

\subsection{Generalization Ability}

Fig. \ref{GeneralRFL}(a) presents the generalization ability for $N_{RF}$ and SNR. We train the DNN in the configuration of $N_{RF} = 12$, $N_{ray}=4$, $L=36$, and SNR$\in \{0, 5, 10 \}$ dB, and test the trained DNN for different values of $N_{RF}$ and SNR with fixed $N_{ray}=4$ and $L=36$. From the figure, we can see that there exists a small performance loss for the DNN employed in different configurations, due to the mismatch of $N_{RF}$ and SNR in the training and testing stages. Moreover, the mismatched DNN still outperforms OPT/OMP and approaches OPT with perfect CSI. This demonstrates the satisfactory generalization ability of the proposed DNN for different values of $N_{RF}$ and SNR. In addition, this illustrates that training the DNN on a wider range of system parameters, e.g., SNR, can help to design more robust systems when perfect prior knowledge about those parameters is unavailable. Furthermore, the performance loss between the mismatched DNN and that without the mismatch decreases with $N_{RF}$ and SNR. This is mainly because there is less performance loss when the mismatch between the training and testing configurations becomes smaller.

Fig. \ref{GeneralRFL}(b) shows the generalization ability for the length of training pilot $L$ and the number of propagating rays $N_{ray}$. We train the DNN in the configuration of $N_{RF}=12$, SNR$= 10$ dB, $L=60$, and $N_{ray}\in \{2, 4, 6 \}$, and test the trained NNs in different settings of $L$ and $N_{ray}$ with fixed $N_{RF}=12$ and SNR$= 10$ dB. It is obvious that the mismatched DNN outperforms OPT/OMP, even though there is a small performance loss compared with the DNN without mismatch. This verifies the satisfactory generalization ability of the proposed DNN for different values of $L$ and $N_{ray}$. Furthermore, it is readily seen that the performance loss decreases with $L$, because when the training pilots are limited, the proposed DNN tends to fully exploit the distribution of the input and adjusts its trainable parameters to fit the particular distribution. In comparison, when the length of training pilot sequences is not the bottleneck, e.g., $L = 64$, the trained DNN can potentially deal with a wider range of channel distributions with different values of $N_{ray}$. 

\begin{figure}[t]
\begin{centering}
\includegraphics[width=0.4\textwidth]{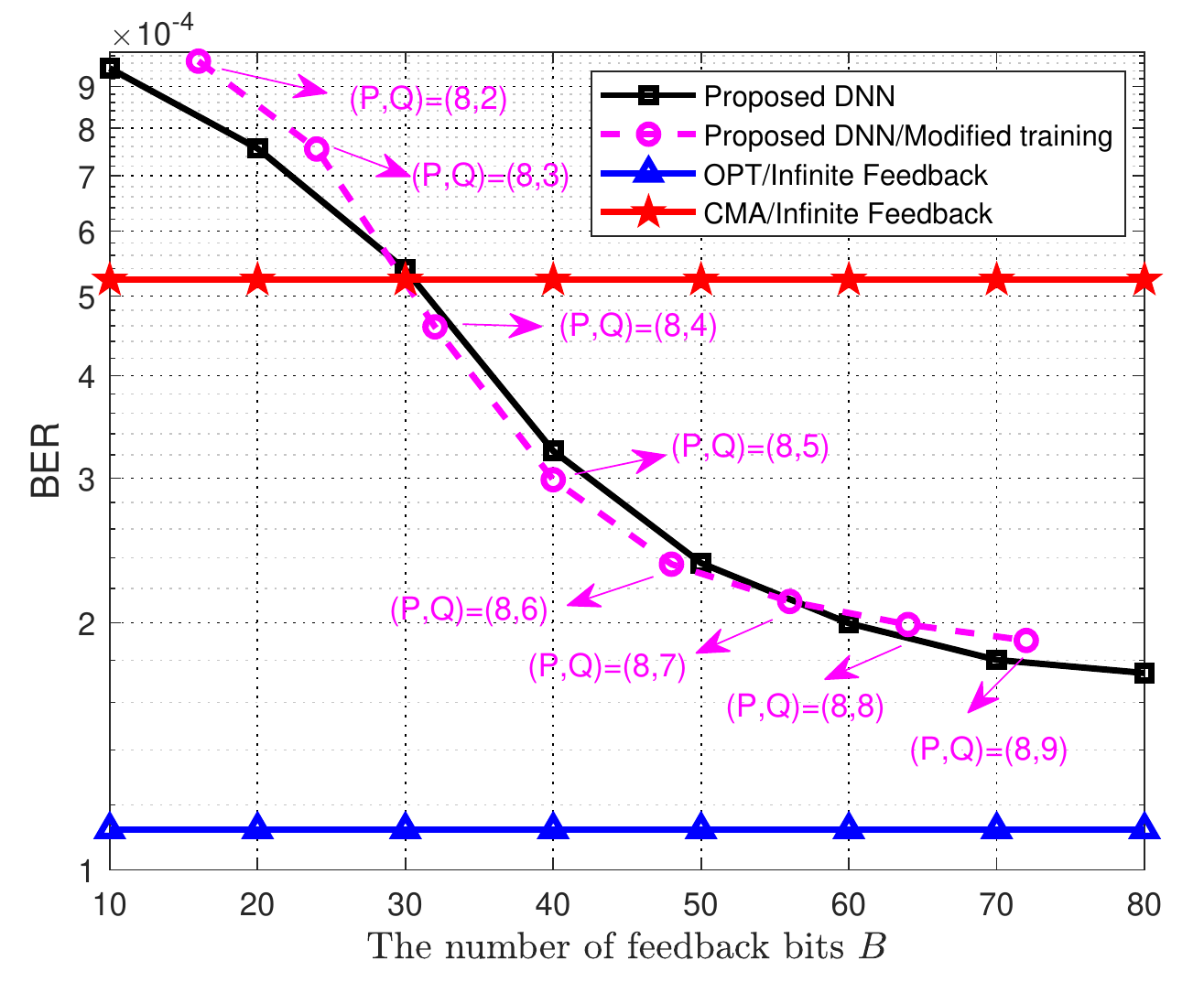}
\par\end{centering}
\caption{Generalization ability for the number of feedback bits $B$ with the modified two-step training method.}
\label{GeneralBit}
\end{figure}

Fig. \ref{GeneralBit} presents the generalization ability for the number of feedback bits $B$. We can see that there is only negligible performance degradation in adopting the modified two-step training method proposed in Section \ref{GeneralB}, which provides a general DNN that can handle different values of $B$. Note that when $B$ is large enough, the proposed DNN achieves the BER performance approaching the lower bound provided by OPT with infinite feedback bits. It shows that the proposed two-step training method can improve the generalization ability of the proposed DNN with respect to $B$, by setting different values of $Q$ in this approach.

\begin{figure}[t]
\begin{centering}
\includegraphics[width=0.4\textwidth]{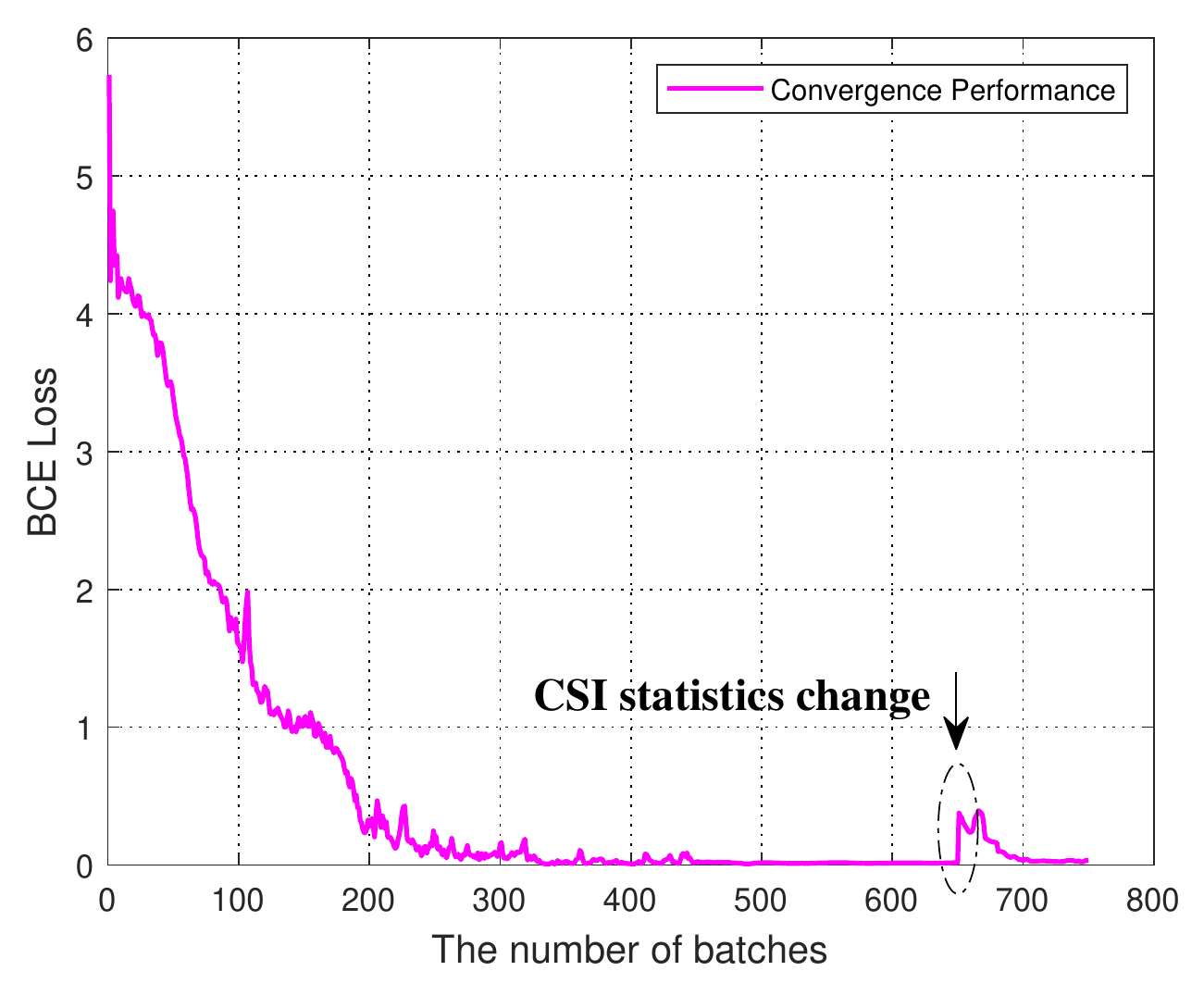}
\par\end{centering}
\caption{Convergence of BCE loss in transfer learning.}
\label{TransferConverge}
\end{figure}

In practice, the channel statistics change continuously and those of adjacent superframes will not change much. Then, the CSI samples from the changed CSI statistics are collected for fine-tuning (online training) based on the previously trained DNN. Specifically, we employ ``transfer learning" to train the DNNs online, where most of the layers of the DNNs are frozen and only the parameters in the last few layers of the DNNs are updated. In this way, the DNNs converge fast and can adapt to the CSI statistics quickly.
Fig. \ref{TransferConverge} presents the BER performance of ``transfer learning". We see that when the CSI statistics change, the BCE loss of DNN increases first, and then decreases within a short time, which shows that the DNN can adapt to the changed CSI statistics quickly.
When the CSI statistics change faster or slower, the frame and time slot length need to be adjusted adaptively. If the CSI statistics change fast, the length of the frame and time slot need to be shortened to obtain more high-dimensional original CSI samples to track the change of CSI, where the long-term analog precoders and combiners are updated more frequently to better fit the change of channel statistics.  

	
\section{Conclusion} \label{Conclusion}
In this paper, we developed a deep learning-based framework for an FDD mmWave massive MIMO system, which consists of DNN-based pilot training, feedback scheme, and hybrid precoding. 
To reduce the heavy signaling overhead and CSI mismatch caused by the delay, a two-timescale DNN composed of a long-term DNN and a short-term DNN has been proposed. 
Furthermore, a two-timescale training method is developed for the proposed DNN with a binary layer.
The proposed two-timescale DNN can be easily extended to OFDM systems.
Simulation results show that our proposed algorithm significantly outperforms conventional schemes in terms of bit-error rate performance with reduced signaling overhead and shorter pilot sequences. Future work includes extending our framework to multi-user and multi-cell systems, and to more challenging problems for future communication systems, such as intelligent reflecting surface systems.


\bibliographystyle{IEEEtran}
\bibliography{IEEEabrv,E2ELearning}

\end{document}